\documentclass[aps,prl,reprint,amsmath,amssymb,superscriptaddress,nolongbibliography,noeprint,nofootinbib,floatfix]{revtex4-2}
\usepackage{graphicx,bm,booktabs,tikz,mathrsfs}
\usetikzlibrary{arrows.meta,calc}
\usepackage[colorlinks=true,citecolor=blue,linkcolor=blue,urlcolor=blue]{hyperref}

\usepackage{xcolor}
\DeclareMathOperator{\Tr}{Tr}

\begin{document}
\AddToHookNext{shipout/foreground}{%
  \put(0,-13mm){%
    \makebox[\paperwidth][r]{%
      \normalfont\small RIKEN-iTHEMS-Report-26\hspace*{18mm}%
    }%
  }%
}
\raggedbottom
\widowpenalty=10000
\title{ De Sitter Gravity and Emergent Time \texorpdfstring{\\}{ }from Fusion Matrix Product States and SYK}
\author{Shinsei Ryu}
\affiliation{Department of Physics, Princeton University, Princeton, New Jersey 08544, USA}
\author{Yu-ki Suzuki}
\affiliation{RIKEN Center for Interdisciplinary Theoretical and Mathematical Sciences (iTHEMS), Wako, Saitama 351-0198, Japan}
\date{\today}

\begin{abstract}
We propose a quantum-mechanical model dual to two-dimensional de Sitter (dS) gravity.
It is a quantum channel, defined by a matrix product state that fuses one spin
at each step, and is related to a limit of the Sachdev-Ye-Kitaev model
without chord crossings.
Its simplest part is a random walk with an emergent $SL(2,\mathbb R)$ symmetry,
which we identify with the symmetry of the future boundary of dS.
It selects a time variable in which the walk obeys the Schwarzian action and
yields the disk amplitude of dS Jackiw-Teitelboim gravity. As the simplest case, we consider free massless scalar fields in the fixed background of dS$_2$. We find the agreement of the two-point function in the Bunch-Davies vacuum and the matching of symmetry. Our construction provides a concrete model for the dS/CFT
correspondence for  two-dimensional de Sitter gravity.
\end{abstract}

\maketitle

\section{Introduction}
The best understood framework for defining quantum gravity is AdS/CFT
\cite{tHooft1993,Susskind1995,Maldacena1998,GubserKlebanovPolyakov1998,Witten1998Holography}.
There, the gravitational theory is defined by an ordinary quantum theory
living on the boundary, that is, a theory with a Hamiltonian generating
time evolution.
A particularly well-understood example is two-dimensional
Jackiw--Teitelboim (JT) gravity~\cite{Teitelboim1983,Jackiw1985},
whose gravitational amplitudes are known for each genus and
number of boundaries.
The amplitudes at all topologies are reproduced by an ensemble
of random matrices~\cite{SaadShenkerStanford2019}, while the low-energy amplitudes are reproduced
by a single quantum mechanical system, the Sachdev--Ye--Kitaev (SYK) model
\cite{SachdevYe1993,Kitaev2015,MaldacenaStanford2016}.

In dS/CFT the situation is different.
The would-be dual theory lives on the spacelike slice at future infinity,
where there is no time
\cite{Witten2001dS,Strominger2001,AnninosHartmanStrominger2011}.
The natural boundary object is therefore expected to be not a Hamiltonian
but a transfer operation that generates the state along the spatial direction.

On the gravitational side, the computable quantity is the path integral:
one fixes conditions on the boundary and sums over all spacetime geometries
satisfying them. Since the boundary of dS lies at future infinity,
this sum assigns an amplitude to each shape of the future boundary;
it is the wavefunction of the universe
\cite{HartleHawking1983,Maldacena2003}.
In general the sum cannot be performed,
but for the dS version of JT gravity (dSJT)
\cite{MaldacenaTuriaciYang2021,CotlerJensenMaloney2020},
Cotler and Jensen carried it out and obtained the amplitudes
with future circles as the boundary
\cite{CotlerJensen2024}.
They observed that these amplitudes behave like those of a matrix integral.
However, no quantum mechanical model reproducing them,
the counterpart of SYK, is known.

The goal of this work is to construct such a model as a quantum channel.
We start from the relation between matrix product states (MPS) and quantum channels.
An MPS on a one-dimensional lattice is generated site by site by a transfer matrix.
Tracing out the physical spins in the transfer matrix gives an evolution
on the internal space, and this evolution is a completely positive,
trace-preserving quantum channel~\cite{FossFeigEtAl2021}.
A quantum channel describes non-unitary evolution.

We construct a specific MPS, which we call the fusion MPS.
At each step it fuses one spin-1/2 into the internal space,
and the simplest part of its channel is a walk in which an integer ``height'' goes
up or down by one at each step.
This is the simplest construction that realizes such a walk
on a Hilbert space with a positive inner product.
As will become clear, the steps of the channel cannot
naively be identified with time evolution in dSJT gravity.

Okuyama gave an MPS representation of the moments of the double-scaled SYK model (DSSYK)
\cite{Okuyama2025TN},
a model whose relation to dS has been discussed~\cite{NarovlanskyVerlinde2023,Okuyama2025dS}.
Our fusion MPS reproduces these moments as well.
We focus on a limiting case of DSSYK, in which the spectral density
becomes a semicircle~\cite{FengTianWei2019}, and call it fusion SYK.
In this limit the walk is especially simple, as noted in
Ref.~\cite{Okuyama2025TN}:
all its paths that return to zero height, known as Dyck paths, have the same weight.
This ensemble of paths has an emergent $SL(2,\mathbb R)$ symmetry in the continuum
\cite{ChenFradkinWitczakKrempa2017}.

The central claim of this work is that this $SL(2,\mathbb R)$
is the symmetry of the future boundary of dS.
It selects a time variable, different from the steps of the channel,
that SYK correlators can measure.
In this time the walk obeys the Schwarzian action,
the same action that governs the boundary of JT gravity.
One caveat is that the fusion MPS lives on a spatial chain separate from future infinity $\mathcal I^+$.
The bulk semi-classical action is fixed separately, by the charge algebra of the fusion MPS:
it is a two-dimensional sinh dilaton gravity that reduces to dSJT gravity in a limit.
Amplitudes of higher topology are not derived in this way.
We reproduce them by a matrix integral, which we take as an input.

Finally, we also derive the dS$_2$ metric from MPS. We confirm the agreement between the Bunch-Davies two-point function of free massless scalar fields in the bulk and overlaps of height basis. Furthermore, the action of bulk isometry is explicitly realized as the MPS operator, which forms the global conformal symmetry.

\section{The fusion MPS}
\label{sec:fusion-chords}
\label{sec:fusion-model}

The internal space of our MPS is $\mathcal H=\bigoplus_{j\in\frac12\mathbb N_0}V_j$, where $V_j$ is the spin-$j$ representation of $U_Q(\mathfrak{su}(2))$, with real $Q\ge1$. Each $V_j$ has an orthonormal basis $|j,m\rangle$, $m=-j,\ldots,j$, so that $\dim V_j=2j+1$.
Each site of the MPS carries a physical spin $1/2$, with basis $|\alpha\rangle$, $\alpha=\pm\tfrac12$.
We define the MPS by a sequential preparation, which can also be read as a quantum channel acting on the internal space~\cite{LeeYoun2022}. At each step, the internal state in some $V_j$ is fused with a spin $1/2$, which is then emitted as the physical spin of that site. The fusion rule allows the internal label to change from $j$ to $J=j\pm\tfrac12$; at $j=0$ only $J=\tfrac12$ occurs. We assign these two branches adjustable probabilities $p_{jJ}\ge0$, with $\sum_Jp_{jJ}=1$ and $p_{0,1/2}=1$; they are the parameters of the construction.
With the $Q$-number $[x]_Q=(Q^x-Q^{-x})/(Q-Q^{-1})$,
and the quantum dimension $d_j=[2j+1]_Q$, the fusion tensors are
\begin{equation}
 (B_\alpha)_{Jm',jm}
 =Q^\alpha\sqrt{\frac{p_{jJ}d_j}{d_J}}
 C_Q(jm;\tfrac12\alpha | Jm').
 \label{eq:prl-fusion-tensors}
\end{equation}
Here $C_Q(jm;\tfrac12\alpha| Jm')$ are the real orthonormal Clebsch--Gordan coefficients of $U_Q(\mathfrak{su}(2))$ for fusing $V_j$ with the spin $1/2$ into $V_J$, in the ordered-coproduct convention of Appendix~\ref{app:mps-definitions}; 
$B_\alpha$ maps $V_j$ to $V_J$. 
One step of the preparation is the map $V_{\rm step}|\psi\rangle=\sum_\alpha B_\alpha|\psi\rangle\otimes|\alpha\rangle$, which is an isometry because $\sum_\alpha B_\alpha^\dagger B_\alpha=I$ (Appendix~\ref{app:mps-definitions})~\cite{SchonEtAl2005}.
Tracing over the emitted spin at each step gives the channel $\widehat{\mathcal E}$~\cite{FannesNachtergaeleWerner1992,CiracEtAl2021}, which maps a density matrix $\rho$ on the internal space $\mathcal H$ to $\widehat{\mathcal E}(\rho)=\sum_\alpha B_\alpha\rho B_\alpha^\dagger$.

The MPS is obtained by preparing the internal state in the trivial representation, $|0\rangle=|0,0\rangle$, applying $N$ steps, and projecting the internal state back onto $|0\rangle$. This final projection, or cap, defines the physical-spin amplitude and its norm,
\begin{equation}
\widetilde\Psi_N(\boldsymbol\alpha)
=\langle0|B_{\alpha_N}\cdots B_{\alpha_1}|0\rangle,\quad
Z_N=\sum_{\boldsymbol\alpha}|\widetilde\Psi_N(\boldsymbol\alpha)|^2 .
\label{eq:proposal-cap-amplitude}
\end{equation}
Since $V_{\rm step}$ is an isometry, $Z_N$ is the probability that the final projection succeeds. For $Z_N>0$, the normalized physical state is $\Psi_N=\widetilde\Psi_N/\sqrt{Z_N}$. 
\footnote{Since the height changes parity at each step, $Z_N$ vanishes for odd $N$.}

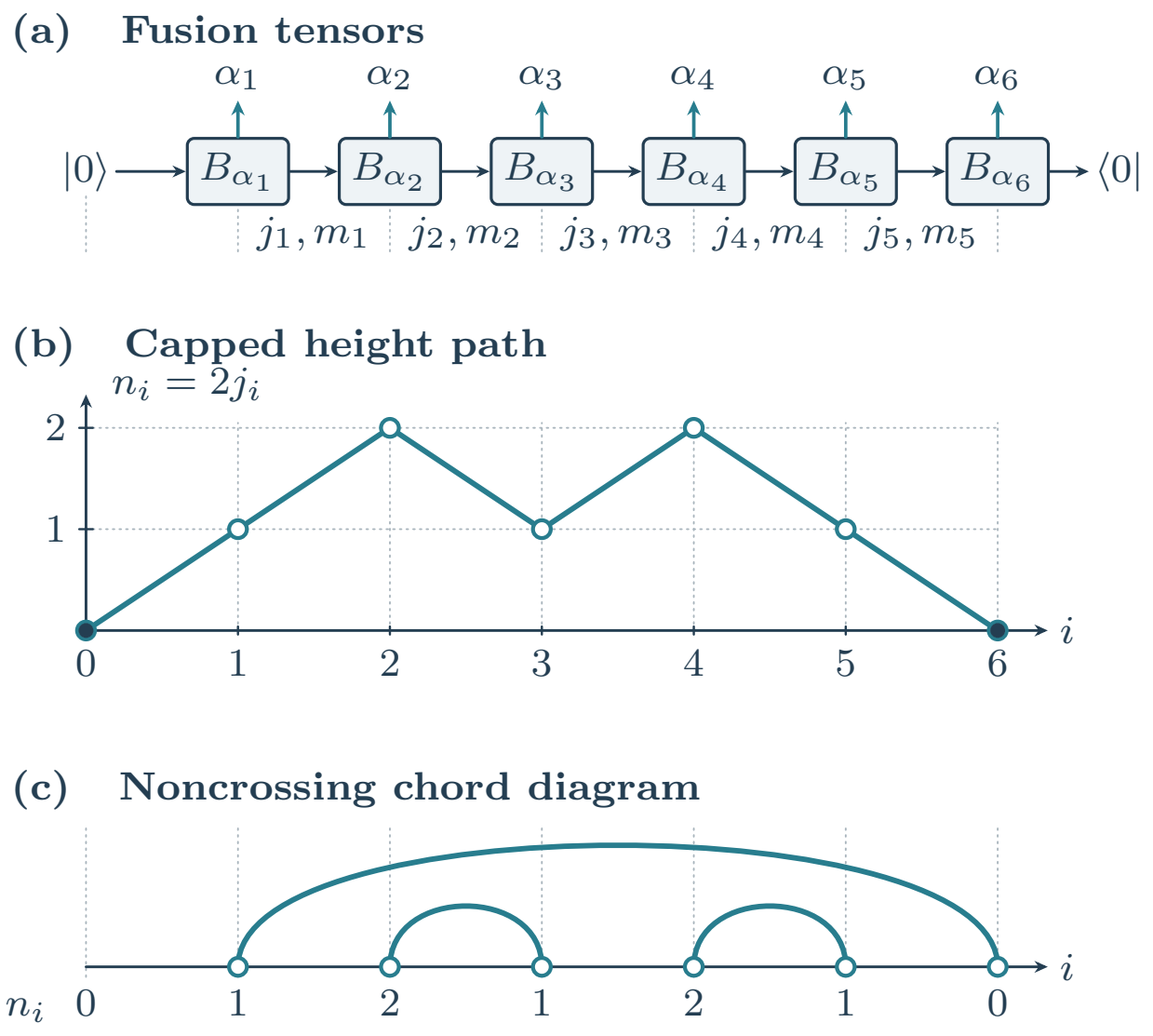
\begin{figure}[!tbp]
\centering
\begingroup
\definecolor{FusionInk}{RGB}{36,62,82}
\definecolor{FusionTeal}{RGB}{40,125,142}
\definecolor{FusionLight}{RGB}{238,243,246}
\definecolor{FusionGuide}{RGB}{169,181,189}
\begin{tikzpicture}[x=1.08cm,y=.72cm,
 every node/.style={font=\footnotesize,text=FusionInk},
 line cap=round,line join=round,
 flow/.style={draw=FusionInk,line width=.55pt,-{Stealth[length=2.7pt,width=2.4pt]}},
 emitted/.style={draw=FusionTeal,line width=.7pt,-{Stealth[length=3pt,width=2.7pt]}},
 step guide/.style={draw=FusionGuide,line width=.35pt,densely dotted},
 tensor/.style={draw=FusionInk,fill=FusionLight,line width=.6pt,
 rounded corners=1.4pt,minimum width=.72cm,minimum height=.47cm,inner sep=1.2pt}]
 \node[anchor=west,font=\footnotesize\bfseries] at (-.60,3.72)
       {(a)\quad Fusion tensors};
 \foreach \i in {0,...,6}
  \draw[step guide] (\i,1.56)--(\i,2.13);
 \foreach \i in {1,...,6}{
  \node[tensor] (b\i) at (\i,2.35) {$B_{\alpha_{\i}}$};
  \draw[emitted] (b\i.north)--(\i,3.05);
  \node[above,inner sep=2pt] at (\i,3.05) {$\alpha_{\i}$};
 }
 \foreach \i in {1,...,5}{
  \pgfmathtruncatemacro{\next}{\i+1}
  \draw[flow] (b\i.east)--(b\next.west);
  \node[below,inner sep=1.5pt] at ({\i+.5},1.99) {$j_{\i},m_{\i}$};
 }
 \node[inner sep=1pt] (vacL) at (0,2.35) {$|0\rangle$};
 \node[anchor=west,inner sep=1pt] (vacR) at (6.60,2.35) {$\langle0|$};
 \draw[flow] (vacL.east)--(b1.west);
 \draw[flow] (b6.east)--(vacR.west);

 \node[anchor=west,font=\footnotesize\bfseries] at (-.60,.62)
       {(b)\quad Capped height path};
 \begin{scope}[yshift=-1.57cm]
  \foreach \i in {1,...,6}
   \draw[step guide] (\i,0)--(\i,2.05);
  \foreach \n in {1,2}{
   \draw[draw=FusionGuide,line width=.35pt,densely dotted]
          (0,\n)--(6,\n);
   \draw[draw=FusionInk,line width=.45pt] (-.045,\n)--(.045,\n);
   \node[left,inner sep=2pt] at (-.06,\n) {$\n$};
  }
  \draw[flow] (0,0)--(6.33,0);
  \draw[flow] (0,0)--(0,2.33);
  \node[anchor=south west,inner sep=1pt] at (.13,2.20) {$n_i=2j_i$};
  \foreach \i in {0,...,6}{
   \draw[draw=FusionInk,line width=.45pt] (\i,-.04)--(\i,.04);
   \node[below,inner sep=2.5pt] at (\i,-.06) {$\i$};
  }
  \draw[draw=FusionTeal,line width=1.1pt]
       (0,0)--(1,1)--(2,2)--(3,1)--(4,2)--(5,1)--(6,0);
  \foreach \x/\y in {0/0,1/1,2/2,3/1,4/2,5/1,6/0}
   \draw[draw=FusionTeal,fill=white,line width=.8pt] (\x,\y) circle (1.8pt);
  \fill[FusionInk] (0,0) circle (1.5pt);
  \fill[FusionInk] (6,0) circle (1.5pt);
  \node[anchor=west,inner sep=2pt] at (6.34,0) {$i$};
 \end{scope}

 \node[anchor=west,font=\footnotesize\bfseries] at (-.60,-3.75)
       {(c)\quad Noncrossing chord diagram};
 \begin{scope}[yshift=-3.96cm]
  \foreach \i/\n in {0/0,1/1,2/2,3/1,4/2,5/1,6/0}{
   \draw[step guide] (\i,-.10)--(\i,1.38);
   \node[below,inner sep=2.5pt] at (\i,-.10) {$\n$};
  }
  \node[anchor=east,inner sep=1pt] at (-.22,-.43) {$n_i$};
  \draw[flow] (0,0)--(6.33,0);
  \node[anchor=west,inner sep=2pt] at (6.34,0) {$i$};
  \draw[draw=FusionTeal,line width=1.1pt]
       (1,0) .. controls (1,1.6) and (6,1.6) .. (6,0);
  \draw[draw=FusionTeal,line width=1.1pt]
       (2,0) .. controls (2,.8) and (3,.8) .. (3,0);
  \draw[draw=FusionTeal,line width=1.1pt]
       (4,0) .. controls (4,.8) and (5,.8) .. (5,0);
  \foreach \i in {1,...,6}
   \draw[draw=FusionTeal,fill=white,line width=.8pt] (\i,0) circle (1.8pt);
 \end{scope}
\end{tikzpicture}
\endgroup
\caption{Fusion tensors, a capped height path, and its chord diagram.
(a) Each $B_{\alpha_i}$ emits a spin and changes $j$ by $\pm\tfrac12$.
(b) The height $n_i=2j_i\ge0$, with $n_0=n_6=0$.
The three panels share the same step positions $i=0,\ldots,6$.
Contracting magnetic indices and summing paths gives the capped MPS.
Tracing the spins gives the height walk.
(c) The same path as a chord diagram of fusion SYK
(Sec.~\ref{sec:syk-moments}): an up step opens a chord and a down step
closes the most recent one, so that $n_i$ counts the chords open
just after step $i$.}
\label{fig:fusion-walk}
\end{figure}

In the contraction of the tensors, each term follows a sequence of internal labels $j_0=0,\,j_1,\,j_2,\ldots$ with $j_{i+1}=j_i\pm\tfrac12$. We call $n_i=2j_i$ the height after step $i$. It is a nonnegative integer that changes by $\pm1$ at each step, and we call the sequence $(n_0,n_1,\ldots)$ a height path. The branch probabilities $p_{jJ}$ weight these paths, whereas the dependence on the magnetic labels $m$ is fixed by the CG coefficients and does not change with $p_{jJ}$. Figure~\ref{fig:fusion-walk} shows the fusion steps and the corresponding height path.

We now read the same construction as a channel. On $V_j$, let $J_z|j,m\rangle=m|j,m\rangle$ and define $G_j=Q^{2J_z}/d_j$, extended by zero outside $V_j$; it has unit trace. The operators $G_j$ span a subspace that the channel maps into itself,
$\widehat{\mathcal E}(G_j)=\sum_Jp_{jJ}G_J$,
which we call the scalar sector. Starting from the vacuum $G_0=|0\rangle\langle0|$, the internal state after $N$ steps, with the emitted spins traced out, therefore stays in this sector,
\begin{equation}
\rho_N=\widehat{\mathcal E}^{N}(G_0)=\sum_{n\ge0}r_n(N)G_{n/2},
\end{equation}
where $r_n(N)\ge0$, $\sum_nr_n(N)=1$, is the probability that the height equals $n$ after $N$ steps. In terms of the height, the branch probabilities read $p_n^\pm=p_{n/2,(n\pm1)/2}$ with $p_0^-=0$, and $r_{N+1}=\mathsf P\,r_N$ with $\mathsf P_{n\pm1,n}=p_n^\pm$. The height path thus becomes a Markov chain on the nonnegative integers, the height walk. Tracing the emitted spins in Eq.~\eqref{eq:proposal-cap-amplitude} gives
$Z_N=\langle0|\widehat{\mathcal E}^{N}(G_0)|0\rangle=r_0(N)=(\mathsf P^N)_{00}$,
so the norm of the unnormalized MPS equals the return probability of the height walk. 


The stochastic matrix $\mathsf P$ is not symmetric. For later purposes it is convenient to use a symmetrized version. Since a closed walk traverses each edge equally often in both directions, the return probabilities depend on $p_n^\pm$ only through the products $p_{n-1}^+p_n^-$, which we may replace by their geometric mean. On $\ell^2(\mathbb N_0)$, with $|n\rangle$ labeling $G_{n/2}$, this defines the symmetric nearest-neighbor (Jacobi) operator~\cite{KarlinMcGregor1959}
\begin{equation}
\begin{aligned}
\mathcal J_a|n\rangle&=a_{n+1}|n+1\rangle+a_n|n-1\rangle,\
\\
a_n^2&=\delta^2p_{n-1}^+p_n^-\quad(n\ge1),\quad a_0=0,
\end{aligned}
\label{eq:prl-scalar-jacobi}
\end{equation}
with $Z_{2k}=\delta^{-2k}\langle0|\mathcal J_a^{2k}|0\rangle$. Here $\delta=Q+Q^{-1}$ is a normalization fixed below. 


Among the branch probabilities, one choice is distinguished by making every height path with the same endpoints equally likely. It is obtained from the quantum dimensions,
\begin{equation}
p_{jJ}=\frac{d_J}{\delta d_j},
\label{eq:proposal-markov-weights}
\end{equation}
which is normalized because $\sum_Jd_J=\delta d_j$. 
We call this choice the Markov-trace reference. These are the branching
weights defined by the Markov trace of the quantum-group fusion
category~\cite{NeshveyevTuset2004}.
Multiplying the branch probabilities along a height path gives $d_{j_N}/(\delta^Nd_{j_0})$: the intermediate quantum dimensions cancel, and every path from the vacuum back to the vacuum has the same probability $\delta^{-N}$. For $N=2k$ the number of such paths is the Catalan number, so
\begin{equation}
Z_{2k}=\frac{\mathrm{Cat}_k}{\delta^{2k}},\quad
\mathrm{Cat}_k=\frac{1}{k+1}\binom{2k}{k}.
\end{equation}
Height paths from the vacuum back to the vacuum are known as Dyck paths.
The same cancellation gives $a_n=1$ in Eq.~\eqref{eq:prl-scalar-jacobi}, so the Jacobi operator reduces to $S+S^\dagger$ with $S|n\rangle=|n+1\rangle$ and $S^\dagger|0\rangle=0$. With an energy scale $E_*>0$ we define the reference Hamiltonian
\begin{equation}
H_0=E_*(S+S^\dagger).
\label{eq:proposal-reference-hamiltonian}
\end{equation}
Its vacuum spectral measure on the height space is the semicircle on $[-2E_*,2E_*]$, with $\langle0|H_0^{2k}|0\rangle=(E_*\delta)^{2k}Z_{2k}=E_*^{2k}\mathrm{Cat}_k$.

More generally, any sequence of Jacobi coefficients can be realized by a suitable choice of branch probabilities. A one-parameter family that will be needed below is $a_n^2=1-q^n$ with $0\le q<1$, obtained from $p_0^-=0$ and the recursion
\begin{equation}
p_n^-=\frac{1-q^n}{\delta^2(1-p_{n-1}^-)},\quad p_n^+=1-p_n^-\quad(n\ge1).
\label{eq:prl-dssyk-weights}
\end{equation}
Since $\delta\ge2$, induction gives $0\le p_n^-\le\tfrac12$, so both branches have well-defined probabilities at every height. At $q=0$ the recursion returns the Markov-trace reference. The meaning of $q$ is explained in the next section.
\section{Hamiltonian moments and SYK realizations}
\label{sec:syk-moments}
The return probabilities are moments of a Hamiltonian on the height space. For $H_{\rm eff}=E_0\mathcal J_a$ with an energy scale $E_0>0$,
\begin{equation}
\langle0|H_{\rm eff}^N|0\rangle=(E_0\delta)^NZ_N .
\label{eq:proposal-hamiltonian-moments}
\end{equation}
The return probabilities at all fusion lengths thus determine the return amplitude $\langle0|e^{-itH_{\rm eff}}|0\rangle$ of the bounded operator $H_{\rm eff}$ (Appendix~\ref{app:reference-returns}). Okuyama obtained the moments of the double-scaled SYK model in this way, by summing physical-spin amplitudes of an MPS~\cite{Okuyama2025TN}; our preparation reproduces them from the branch probabilities while retaining the magnetic and nonscalar sectors.
The SYK model consists of an even number $N_f$ of Majorana operators, $\{\chi_i,\chi_j\}=2\delta_{ij}$, with the Hamiltonian~\cite{MaldacenaStanford2016}
\begin{equation}
H_{\rm SYK}=\frac{E_*}{\sqrt{\binom{N_f}{p}}}
\sum_{|I|=p}J_I i^{p/2}\prod_{i\in I}\chi_i ,
\label{eq:proposal-microscopic-syk}
\end{equation}
where $p$ is even, the product is in increasing index order, and the couplings $J_I$ are independent standard Gaussian variables drawn once per sample. With the normalized trace $\operatorname{tr}=2^{-N_f/2}\Tr$ and the disorder average $\mathbb E$, one has $\mathbb E\operatorname{tr}(H_{\rm SYK}^2)=E_*^2$. (We use $N_f$ for the Majorana count and $N$ for the fusion length.)
Disorder averaging pairs the interaction terms~\cite{GarciaGarciaJiaVerbaarschot2018}. In the double-scaled limit $N_f,p\to\infty$ with $2p^2/N_f\to-\log q$, $0<q<1$, each pairing is drawn as a chord diagram and contributes $q^{\rm cr}$, with ${\rm cr}$ the number of crossings~\cite{BINT2019,Lin2022,CotlerEtAl2017}:
\begin{equation}
\lim_{\rm DS}\frac{\mathbb E\operatorname{tr}(H_{\rm SYK}^{2k})}{E_*^{2k}}
=m_{2k}(q)=\sum_{\pi\in\mathcal P_2(2k)}q^{{\rm cr}(\pi)} .
\label{eq:proposal-syk-chord-moments}
\end{equation}
Reading the chord endpoints in order gives a walk whose height counts the open chords [Fig.~\ref{fig:fusion-walk}(c)], and closing a chord at height $n$ carries the weight $1+q+\cdots+q^{n-1}=[n]_q$~\cite{IsmailStantonViennot1987,RabinoviciEtAl2023}. This is the family of Eq.~\eqref{eq:prl-dssyk-weights}: with $H_q=E_*(1-q)^{-1/2}\mathcal J_a$ one finds $\langle0|H_q^{2k}|0\rangle=E_*^{2k}m_{2k}(q)$. The crossing parameter $q$ is unrelated to the fusion parameter $Q$; at fixed $q$, changing $Q$ changes the magnetic data and leaves the moments unchanged. The DSSYK chord algebra admits quantum-group descriptions based on a deformation of $SU(1,1)$~\cite{BerkoozIsachenkovNarayanNarovlansky2023,BlommaertMertensYao2024Actions}.

In this paper we focus on the endpoint $q=0$, where only noncrossing chord diagrams survive, $m_{2k}(0)={\rm Cat}_k$, and $H_q$ reduces to the reference Hamiltonian $H_0$~\cite{Okuyama2025TN}. Within the same Gaussian family, this endpoint is reached at fixed $k$ in the limit
\begin{equation}
\frac{p^2}{N_f}\to\infty,\qquad \frac{p^3}{N_f^2}\to0,
\label{eq:proposal-markov-microscopic-limit}
\end{equation}
for instance $p\asymp N_f^{3/5}$. For even $p\le N_f/2$, the first condition suppresses crossed pairings and gives the semicircle law~\cite{FengTianWei2019}. We use the second condition to control higher overlaps in the interpolation. We call this the noncrossing limit of SYK and the resulting model fusion SYK; its height walk is the Markov-trace reference of Sec.~\ref{sec:fusion-model}. A continuous interpolation between a prescribed $q_*$ and the noncrossing endpoint, at fixed Majorana Hilbert space and fixed couplings, is given in Appendix~\ref{supp:syk-interpolation}.

The moments at $q=0$ do not depend on $Q$, but the response to a pair of probes does. Take an independent Gaussian Majorana sum $O$ of even order $p_{\rm pr}$ with $\mathbb E\operatorname{tr}(O^2)=1$, scaled so that $pp_{\rm pr}/N_f\to\log Q$. With the height operator $\hat n|n\rangle=n|n\rangle$, the matter-chord rule gives~\cite{BINT2019,LiuShen2026}
\begin{equation}
\mathbb E\operatorname{tr}(H_{\rm SYK}^{k_1}OH_{\rm SYK}^{k_2}O)
\longrightarrow\langle0|H_0^{k_1}Q^{-2\hat n}H_0^{k_2}|0\rangle .
\label{eq:proposal-paired-probe}
\end{equation}
The factor $Q^{-2\hat n}$ arises from the chord connecting the two insertions of $O$, which crosses every chord open between them. It is this paired response that we use below to read the height from fermion correlators.
The nonscalar sectors of the reference channel have their own transfer spectrum; Appendix~\ref{app:transfer-resonances} analyzes its large-$Q$ pole pattern.

\section{Gravity dual}
\label{prl:sec:gravinterpretation}
\label{sec:gravity-2d}

\emph{Classical action.---}
We now identify the action of the gravity dual by asking which classical gravity has the charge algebra of the fusion MPS. Let $h_g>0$ be a small parameter, to be identified with the gravitational coupling below, and define $\widehat\phi=h_gJ_z$, $\widehat Y_1=h_g(E+F)/2$, and $\widehat Y_2=h_g(E-F)/(2i)$, with $\log Q=\lambda h_g/2$. The quantum-group relations give $[\widehat\phi,\widehat Y_1]=ih_g\widehat Y_2$, $[\widehat\phi,\widehat Y_2]=-ih_g\widehat Y_1$, and $[\widehat Y_1,\widehat Y_2]=ih_g\gamma\sinh(\lambda\widehat\phi)/(\lambda\sinh\gamma)$ with $\gamma=\log Q$. In the limit $h_g\to0$ with $h_gm$, $h_g(j+\tfrac12)$, and $\lambda$ fixed, the brackets $\{f,g\}=\lim[\widehat f,\widehat g]/(ih_g)$ become
$\{\phi,Y_1\}=Y_2$, $\{\phi,Y_2\}=-Y_1$, and $\{Y_1,Y_2\}=V_\lambda(\phi)=\sinh(\lambda\phi)/\lambda$.
This limit sends $Q\to1$ at fixed $\lambda=2\log Q/h_g$.

In the Poisson sigma model with target $(\phi,Y_1,Y_2)$~\cite{SchallerStrobl1994,Ikeda1994},
the scalars $Y_a$ impose vanishing torsion. With frame $A_{Y_a}=e^a$
and dilaton $\phi$ (Appendix~\ref{app:psm}), the local action is
\begin{equation}
 I_E=-\frac{A_g}2\int_\Sigma\sqrt g\,
       [\phi R-2V_\lambda(\phi)]\,d^2x+I_\partial ,\quad A_g=\frac{\hbar}{h_g}.
\label{prl:eq:sinhaction}
\end{equation}
The coefficient $A_g$ plays the role of the inverse Newton constant, so $h_g$ counts loops, and the classical limit above is the weak-gravity limit $A_g\to\infty$ at fixed $\lambda=2A_g\log Q/\hbar$. The bulk action is thus fixed by the charge algebra up to the normalization $A_g$; the hyperbolic potential follows from the reality of $Q$, a phase $Q$ giving sine dilaton gravity instead~\cite{GrumillerKummerVassilevich2002}. A sine-dilaton description of DSSYK was obtained in Ref.~\cite{BlommaertMertensPapalini2025}. The $q$-Schwarzian description of the $\sinh$ dilaton disk uses a noncompact quantum group~\cite{BlommaertMertensYao2024Liouville}. 
Here $I_\partial$ is the appropriate boundary term, and the further limit $\lambda\to0$ gives dSJT gravity.
The identification of channel amplitudes with gravitational amplitudes is not derived here; it is the subject of the remainder of the section.

\emph{dSJT amplitudes.---}
To test the identification, we need the gravitational amplitudes at $\lambda=0$, that is, of dSJT gravity, with one or more future circles as the boundary. Cotler and Jensen (CJ)~\cite{CotlerJensen2024} computed them by integrating over complex metrics, so that the geometries can be described by hyperbolic surfaces. Each future circle is attached to the rest of the surface through a trumpet, the region between the circle and a closed geodesic. The boundary condition on a future circle fixes the regulated proper length $\ell_b$ and the boundary value $\varphi_b$ of the normalized dilaton $\varphi=A_g\phi/2$. As in AdSJT, where the circle enters only through the inverse temperature $\beta\propto\ell_b/\varphi_b$, here it enters through a single complex parameter
\begin{equation}
\tau_F=\frac{1}{0^+-i\Phi_F},\qquad \Phi_F=\frac{4\pi^2\varphi_b}{\ell_b}.
\label{eq:tauF}
\end{equation}
In CJ's conventions, the continued parameter is
$\beta_{\rm CJ}=-\tau_F$. The $0^+$ regulates the trumpet integral.
The geodesic has a continued length $b=2\pi i\sqrt P$, with $P\ge0$, and the trumpet contributes
\begin{equation}
T_{\tau_F}(P)=\frac{e^{-P/\tau_F}}{\sqrt{2\pi\tau_F}},
\label{eq:trumpet}
\end{equation}
which is integrable over $P$ when $\operatorname{Im}\Phi_F>0$; with several future circles, each circle $i$ has its own $\tau_{F,i}$ and $P_i$. 
The Euler term contributes $e^{S_0\chi}$, with
$\chi=2-2g-n_{\rm b}$ and $g_s=e^{-S_0}$.
The disk cap at $P=1$ gives
\begin{equation}
Z_{0,1}=g_s^{-1}\int dP\,\delta'(P-1)\,T_{\tau_F}(P)
=\frac{e^{-1/\tau_F}}{\sqrt{2\pi}\,g_s\,\tau_F^{3/2}} .
\label{eq:cj-disk}
\end{equation}
For $n_{\rm b}\ge1$ and $2g-2+n_{\rm b}>0$, we sew the trumpets to a
genus-$g$ core and integrate over its moduli.
Appendix~\ref{app:seeds} gives the sewing formula and the disk and
cylinder seeds. Figure~\ref{fig:gravity-sewing} illustrates the gluing.

\emph{Matrix realization.---}
As in AdSJT~\cite{SaadShenkerStanford2019}, we use the genus
expansion of a matrix integral. The CJ amplitudes follow by analytic
continuation, and the matrix potential is expressed by the return amplitude
of the Markov-trace reference,
$\mathcal R(z)=\langle0|e^{zW}|0\rangle=I_1(2z)/z$ with $W=H_0/E_*=S+S^\dagger$,
which does not depend on $Q$. The matrix integral itself is an input of the construction and is not derived from the channel; Appendix~\ref{supp:matrix-preparation} gives a collective preparation of its measure and its genus expansion.

Consider an $M\times M$ Hermitian matrix with eigenvalues $x_a$ and the integral
\begin{equation}
 \mathcal Z_M=\int_{\mathcal C_\epsilon^M}\prod_{a=1}^Mdx_a\,\Delta(x)^2
 \exp\Big[-\frac1{\hbar_m}\sum_aV_\epsilon(x_a)\Big],
\end{equation}
where $\Delta(x)=\prod_{a<b}(x_a-x_b)$ and $V_\epsilon$ is an even potential; the parameter $\epsilon$ sets the energy scale below. 
The real integration interval $\mathcal C_\epsilon$ encloses the cut (Appendix~\ref{supp:matrix-preparation}). In the planar limit, $M\to\infty$ at fixed $t=M\hbar_m$, the eigenvalues fill an interval $[-2\sqrt r,2\sqrt r]$, and the saddle-point equation relates $t$ to $r$: for
$V_\epsilon(x)=\sum_{j\ge1}A_jx^{2j}/[2j\binom{2j-1}{j}]$
one has $t(r)=\sum_{j\ge1}A_jr^j$. We fix the coefficients $A_j$ by requiring
\begin{equation}
 t(1)-t(r)=\frac{1-r}{2}\,
       \mathcal R\!\left(\frac{\pi\sqrt{1-r}}\epsilon\right).
\label{eq:gravity-matrix-weight}
\end{equation}
With $u=(1-r)/\epsilon^2$ the right-hand side is $\epsilon^2\sqrt u\,I_1(2\pi\sqrt u)/(2\pi)$, which is the genus-zero string equation of JT gravity~\cite{OkuyamaSakai2020}: the return amplitude of the channel and the JT string equation are the same function. At $r=1$ the eigenvalue density, normalized to $\int\rho\,dx=t(1)$, is then exactly
\begin{equation}
 \rho(x)=\frac{\epsilon}{4\pi^2}\sinh\Big(\frac{\pi}{\epsilon}\sqrt{4-x^2}\Big),\quad -2\le x\le2 .
\label{eq:gravity-edge-density}
\end{equation}
We zoom into the upper edge, $x=2-\epsilon^2E$, and take $\epsilon\to0$ at fixed $\kappa=\hbar_m/\epsilon^3$. The edge density before continuation is $\rho_{\rm edge}(E)=\sinh(2\pi\sqrt E)/(4\pi^2\kappa)$, and the disk amplitude is
\begin{equation}
 D(\beta)=\int_0^\infty \rho_{\rm edge}(E)e^{-\beta E}dE
 =\frac{e^{\pi^2/\beta}}{4\sqrt\pi\kappa\beta^{3/2}},
 \quad\operatorname{Re}\beta>0.
\label{eq:gravity-disk-dictionary}
\end{equation}
With $\beta=-\pi^2\tau_F$, $\sqrt{\beta}=-i\pi\sqrt{\tau_F}$,
and $\kappa=-i\alpha_{\rm JT}^{3/2}g_s$, where
$\alpha_{\rm JT}=(2\pi^2)^{-1}$, we obtain
$D=Z_{0,1}$ in Eq.~\eqref{eq:cj-disk}. On gravity side, Cotler and Jensen showed that the dSJT genus expansion follows from
the AdSJT matrix-integral expansion by analytic continuation of the
boundary parameters and the string coupling~\cite{CotlerJensen2024}.
On the matrix theory side, the amplitudes of higher genus and with more boundaries follow from the loop equations~\cite{EynardOrantinTopRec2007,EynardOrantin2007,SaadShenkerStanford2019}, and agree with the CJ amplitudes of Appendix~\ref{app:seeds} at each fixed genus and boundary number. A related dS construction uses the upper edge of DSSYK~\cite{Okuyama2025dS}. Distinct nonperturbative completions can share the same genus expansion,
as illustrated in AdS JT gravity~\cite{Johnson2020}.

\emph{Schwarzian.---}
In the matrix realization, the $\sinh$ density was an input. We now obtain the disk from the height walk of Sec.~\ref{sec:fusion-model}, the random walk performed by the height $n$ along the chain. In the Markov-trace reference it is the equal-weight ensemble of Dyck paths. This ensemble, which also describes the ground state of the Fredkin chain~\cite{SalbergerKorepin2016}, has an emergent conformal symmetry in the continuum~\cite{ChenFradkinWitczakKrempa2017}. We show that this symmetry selects a natural time variable, that fusion SYK correlators measure it, and that the height walk written in this time is the Schwarzian theory of the dSJT disk. Details are given in Appendix~\ref{app:reparametrization}.

We start from the continuum limit. Since all height paths of a given length have the same weight, the walk is Brownian motion for long paths at large heights. A height path, now a function $n(T)$, carries the weight $e^{-{\cal S}}$ with
\begin{equation}
{\cal S}=\frac14\int dT\,(\partial_Tn)^2 ,
\label{eq:S0}
\end{equation}
where $T$ is the coordinate along the chain, the continuous counterpart of the step number. This is the action of a free particle with position $n$ and Euclidean time $T$.

This action has a conformal symmetry~\cite{ChenFradkinWitczakKrempa2017}. Up to endpoint terms, ${\cal S}$ is invariant under
\begin{equation}
T\to\frac{aT+b}{cT+d},\qquad n\to\frac{n}{cT+d},
\label{eq:mobius}
\end{equation}
with real $a,b,c,d$ and $ad-bc=1$. This $SL(2,\mathbb R)$ relates different height paths of the same weight. It is not a symmetry of a single step of the channel.

The symmetry selects a natural time variable. The coordinate $T$ is not invariant under Eq.~\eqref{eq:mobius}, but the combination
\begin{equation}
d\sigma=\frac{dT}{n^2}
\label{eq:invariant-time}
\end{equation}
is. We call $\sigma$ the invariant time
or "clock"; it plays the role that proper time plays in relativity. Its total amount along a height path is $I=\int d\sigma$. 
We evaluate it on an interior interval with positive endpoint heights. 
The fusion SYK measures the invariant time: the paired probes of Eq.~\eqref{eq:proposal-paired-probe} insert $e^{-w\hat n}$ with $w=2pp_{\rm pr}/N_f$, and $\int_0^\infty we^{-w}e^{-w\hat n}dw=(\hat n+1)^{-2}$, which is $n^{-2}$ at large heights.

In the invariant time the walk becomes a Schwarzian theory. With $x=\log n$, Eq.~\eqref{eq:S0} reads ${\cal S}=\frac14\int_0^Id\sigma\,(\partial_\sigma x)^2$, so the log-height is a free particle in the invariant time. Summing $e^{-{\cal S}}$ over all height paths of total invariant time $I$ along which $x$ changes by $\ell$ gives the propagator of this free particle. Up to endpoint factors, it is
\begin{equation}
A_0(\ell,I)=\frac{e^{-I/4-\ell^2/(4I)}}{\sqrt{4\pi I}},
\label{eq:free-propagator}
\end{equation}
where $e^{-I/4}$ comes from the change of variables in the measure. A height path can equally be specified by the chain coordinate as a function of the invariant time, $\zeta(u)=T(\sigma)$ with $u=\sigma/I$. Since $\zeta'\propto e^{2x}$,
\begin{equation}
{\cal S}=-\frac1{8I}\int_0^1du\,\{\zeta,u\}+\frac{x'(1)-x'(0)}{4I},
\label{eq:Sx-schwarzian}
\end{equation}
with $\{\zeta,u\}=\zeta'''/\zeta'-\tfrac32(\zeta''/\zeta')^2$. The Schwarzian derivative appears because it is the simplest quantity invariant under M\"obius transformations of $\zeta$, which is the symmetry~\eqref{eq:mobius}.

\begin{figure}[!tbp]
\centering
\begingroup
\definecolor{gravityInk}{RGB}{36,62,82}
\definecolor{gravityTeal}{RGB}{40,125,142}
\definecolor{gravityLight}{RGB}{238,243,246}
\definecolor{gravityMuted}{RGB}{111,122,131}
\begin{tikzpicture}[x=1cm,y=1cm,scale=.90,
 every node/.style={font=\fontsize{8}{9.2}\selectfont,text=gravityInk},
 draw=gravityInk,line width=.65pt,line cap=round,line join=round]
 \path[use as bounding box] (0,-.55) rectangle (7.9,3.75);

 \node[anchor=west,font=\footnotesize\bfseries] at (0,3.55) {(a)\quad Trumpet};
 \begin{scope}[xshift=1.75cm]
  \path[fill=gravityLight,draw=none]
   (-.30,.70)
   .. controls (-.34,1.18) and (-.78,2.32) .. (-.95,2.75)
   -- (.95,2.75)
   .. controls (.78,2.32) and (.34,1.18) .. (.30,.70)
   -- cycle;
  \draw (-.30,.70) .. controls (-.34,1.18) and (-.78,2.32) .. (-.95,2.75);
  \draw (.30,.70) .. controls (.34,1.18) and (.78,2.32) .. (.95,2.75);
  \draw[fill=white] (0,2.75) ellipse (.95 and .13);
  \draw[fill=white,draw=gravityTeal,line width=.85pt]
   (0,.70) ellipse (.30 and .08);
  \node at (0,3.10) {future circle, $\tau_F$};
  \node at (0,2.10) {$T_{\tau_F}(P)$};
  \node[text=gravityTeal] at (0,.40) {geodesic, $P$};
 \end{scope}

 \draw[-{Stealth[length=2.6pt,width=2.4pt]},draw=gravityMuted]
  (3.20,1.85)--(4.55,1.85);
 \node[text=gravityMuted] at (3.875,2.10) {$-\partial_P$};
 \node[text=gravityMuted] at (3.875,1.58) {at $P=1$};

 \node[anchor=west,font=\footnotesize\bfseries] at (4.55,3.55) {(b)\quad Disk};
 \begin{scope}[xshift=6.05cm]
  \path[fill=gravityLight,draw=gravityInk]
   (-.95,2.75)
   .. controls (-.78,2.32) and (-.36,1.50) .. (-.30,1.12)
   .. controls (-.25,.72) and (.25,.72) .. (.30,1.12)
   .. controls (.36,1.50) and (.78,2.32) .. (.95,2.75);
  \draw[fill=white] (0,2.75) ellipse (.95 and .13);
  \fill[gravityTeal] (0,.82) circle (1.5pt);
  \node at (0,3.10) {future circle, $\tau_F$};
  \node at (0,2.10) {$Z_{0,1}$};
  \node[text=gravityTeal] at (0,.40) {$P=1$};
 \end{scope}

 \node at (3.95,-.25)
  {$T_{\tau_F}(P)\leftrightarrow A_0(\ell,I)$,\quad
   $4I=-\pi^2\tau_F$,\quad $P=-\ell^2/\pi^2$};
\end{tikzpicture}
\endgroup
\caption{Trumpet and disk.
(a) The trumpet connects a future circle with parameter $\tau_F$ to a
closed geodesic labeled by $P$ and contributes $T_{\tau_F}(P)$. It
corresponds to the propagator $A_0(\ell,I)$ of the log-height.
(b) The disk is the trumpet closed at $P=1$, where the geodesic shrinks to
a point; it is obtained by applying the seed $-\partial_P$ at $P=1$.
The general gluing is shown in Fig.~\ref{fig:gravity-sewing}.}
\label{fig:trumpet-disk}
\end{figure}
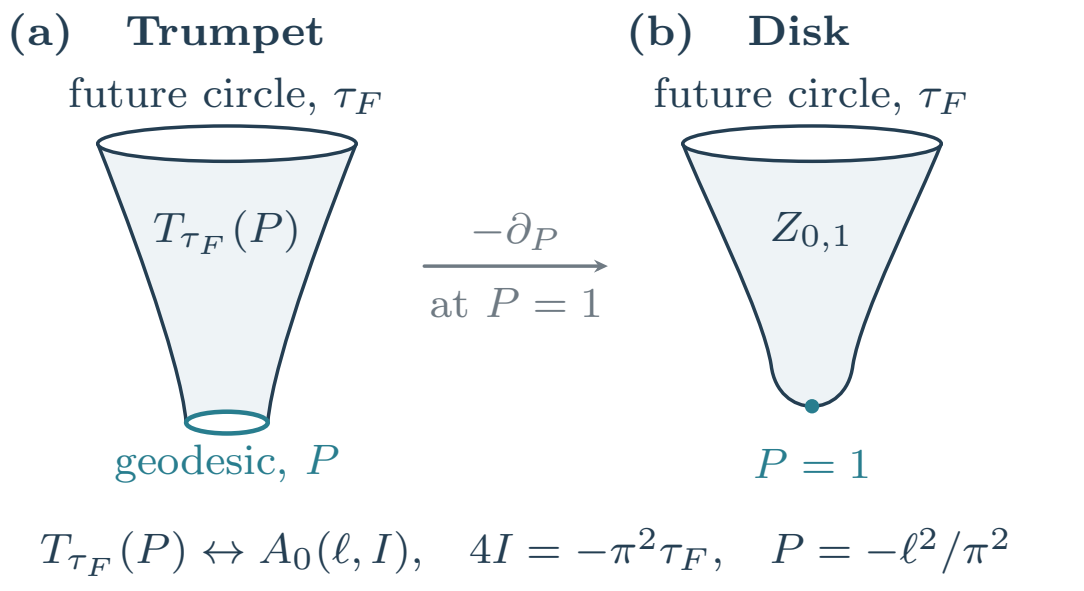

Finally, we take the invariant time to be periodic, so that $u$ is a coordinate on a circle. In gravity this is the future circle. A reparametrization $f$ of the circle, $f(u+1)=f(u)+1$, corresponds to $\zeta_f=C_0+C_1e^{2\pi if(u)}$, an analytic continuation of the real height paths. Then $x$ changes by $\ell=i\pi$ in one circuit, the endpoint term vanishes, and
\begin{equation}
{\cal S}=-\frac1{8I}\int_0^1du\,\{\tan\pi f(u),u\},
\label{eq:Sch-circle}
\end{equation}
the Schwarzian action of the boundary of JT gravity and of low-energy SYK~\cite{MaldacenaStanfordYang2016,MaldacenaStanford2016}. With $P=-\ell^2/\pi^2$ the propagator is $A_0\propto e^{\pi^2P/(4I)}$. This is the trumpet of Eq.~\eqref{eq:trumpet} with $4I=-\pi^2\tau_F$ [Fig.~\ref{fig:trumpet-disk}(a)], and the circle is $P=1$. Applying the disk seed, $-\partial_P$ at $P=1$, reproduces the CJ disk of Eq.~\eqref{eq:cj-disk} up to a constant and the factor $e^{-I/4}$ [Fig.~\ref{fig:trumpet-disk}(b)].   Removing the factor $e^{-I/4}$ sets the energy origin at the threshold
of the clock Hamiltonian. The $\sinh$ density, an input of the matrix realization, is here a consequence. The gravitational boundary condition sets the total invariant time to the imaginary value $I_F=-i\pi^2/(4\Phi_F)$. The action of $PSL(2,\mathbb R)$ and of general reparametrizations on the height states of the MPS is described in Appendix~\ref{app:reparametrization}.

\emph{Metric of $\mathrm{dS}_2$.---}
We reconstruct a local bulk metric from the bilocal of a single height path.

In $\mathrm{dS}_2$, the two light rays emitted from a bulk point reach the future boundary at two points, $u<v$. The pair $(u,v)$ labels the bulk point, $u$ and $v$ are null coordinates, and the metric is $ds^2=\Omega(u,v)\,du\,dv$ with a single unknown function $\Omega$.

In the height walk, we take $u$ and $v$ to be two values of the invariant time along a height path, in units of $I$. We use the projectively invariant bilocal~\cite{MertensTuriaciVerlinde2017,Losev2024}
\begin{equation}
\mathcal O(u,v)=\frac{\sqrt{\zeta'(u)\zeta'(v)}}{\zeta(v)-\zeta(u)}\propto\frac{n_un_v}{T_v-T_u},
\label{eq:bilocal}
\end{equation}
the product of the two heights divided by their distance along the chain. It is invariant under Eq.~\eqref{eq:mobius}.

The area of the small diamond between $(u,v)$ and $(u+\delta u,v+\delta v)$, proportional to $|\Omega|\,\delta u\,\delta v$, depends only on its four boundary points and not on their parametrization. The combination of $\mathcal O$ with this property is
\begin{equation}
\mathcal C=\frac{\mathcal O(u+\delta u,v)\,\mathcal O(u,v+\delta v)}{\mathcal O(u+\delta u,v+\delta v)\,\mathcal O(u,v)},
\label{eq:cross-ratio}
\end{equation}
in which the factors of $\zeta'$ cancel, leaving the cross ratio of the four values of $\zeta$ (cf. the kinematic-space construction of Ref.~\cite{CzechLamprouMcCandlishSully2015}). Since $\log\mathcal C=\mathcal O^2\delta u\,\delta v$ for a small diamond, we identify $\Omega$ with $\mathcal O^2$ up to a constant,
\begin{equation}
ds^2=4\,\mathcal O(u,v)^2\,du\,dv=\frac{4\,dz_u\,dz_v}{(z_v-z_u)^2},
\label{eq:ds2-metric}
\end{equation}
where $z_u=\zeta(u)$ and $z_v=\zeta(v)$. 

In the static patch $z_u<0<z_v$, setting $z_u=-e^{t_{\rm st}-r_*}$, $z_v=e^{t_{\rm st}+r_*}$, and $r=\tanh r_*$ gives
\begin{equation}
ds^2=-(1-r^2)\,dt_{\rm st}^2+\frac{dr^2}{1-r^2},\qquad|r|<1,
\label{eq:static-patch}
\end{equation}
where we choose the sign for which the observer's static Killing field is
timelike in signature $(-,+)$ and the resulting metric has $R=2$.
A common rescaling of $z_u$ and $z_v$, one of the transformations~\eqref{eq:mobius}, shifts the static time $t_{\rm st}$.

\emph{Emergence of time and conformal symmetry.---}
On the virtual height space, the bulk shift
$t_{\rm st}\mapsto t_{\rm st}+s$ at fixed $r$ is represented by
$U_{\rm st}(s)=e^{-isB_{\rm st}}$, where
$B_{\rm st}=i(J_+-J_-)/2$ (Appendix~\ref{app:static-height-evolution}).
For a free massless scalar $\phi$ on fixed $\mathrm{dS}_2$, height-mode
overlaps reproduce the Bunch--Davies Wightman kernels of the weight-one
currents $j_\pm=\partial_{\xi_\pm}\phi$, where
$\xi_\pm=t_{\rm st}\pm r_*$. The holomorphic and antiholomorphic kernels
have the same form in their respective null coordinates. The mixed bulk
correlators vanish, so the sum of the two kernels gives the Wightman
function of $\partial_{t_{\rm st}}\phi=j_++j_-$
(Appendix~\ref{app:height-bd-wightman}).

\section{Conclusion}
\label{sec:conclusion}

In general relativity, time is a coordinate on the same footing as space.
How time arises in quantum gravity is not well understood, and the question is
sharpest in de Sitter space. In AdS/CFT the boundary theory has its own
time, and what emerges is the radial direction of space. In dS/CFT the
dual theory lives at future infinity, which is spacelike. It has no time
of its own, and the time of gravity must emerge from a purely spatial
quantum system.
The fusion MPS gives a concrete example of this emergence. Its chain is
spatial, and the steps of its channel are not the time of gravity.
Instead, the height of the walk sets the rate of the invariant time, the
time variable selected by the emergent $SL(2,\mathbb R)$ symmetry. This
time is not a parameter of the model. It is accumulated along each height
path, as proper time is accumulated along a worldline, and the height that
sets its rate is the total spin of all the spins emitted so far, not a
local degree of freedom of the chain. In this time the walk obeys the
Schwarzian action and gives the disk amplitude of Cotler and Jensen. 
Pairs of points along the walk, read as the endpoints of light rays,
determine the metric of dS$_2$, and the isometries of dS$_2$, including
time translation in the static patch, act on the internal space by mixing
the heights. In MPS, this can be realized as the global conformal transformations acting on the height basis.
The charge algebra of the MPS fixes
the semi-classical bulk action. The dual of dS is expected to be nonunitary; here the
internal space keeps a positive inner product, and the complex phases of
the dS amplitudes arise from the analytic continuation of the invariant
time. Thus, our construction provides an explicit holographic dictionary relating
observables in two-dimensional de Sitter gravity to a quantum-mechanical
model defined by a fusion MPS. Whether the same mechanism can account for the emergence of time in
full quantum gravity of our universe remains open.


Several steps remain open. The amplitudes of higher topology were reproduced by a matrix integral whose potential is designed rather than derived; deriving it from the MPS is the main open problem. The genus expansion leaves a choice of nonperturbative completion, as illustrated in AdS JT gravity~\cite{SaadShenkerStanford2019,Johnson2020}. The symmetry relies on the equal weights of the noncrossing limit, and it is not known whether it survives in double-scaled SYK. The role of $Q$ in the amplitudes also remains to be understood.
The fusion MPS contains more than the height walk. Its magnetic sector depends on $Q$ and determines the entanglement spectrum of the physical spins and the transfer resonances outside the scalar sector (Appendix~\ref{supp:entanglement}). 
At $Q=1$ the Markov-reference tensors admit a chiral Wess--Zumino--Witten lift~\cite{CiracSierra2010,NielsenCiracSierra2011,FrancoRubioCiracSierra2025} (Appendix~\ref{supp:wzw}). Height-dependent filters give conformal and Liouville mechanics~\cite{deAlfaroFubiniFurlan1976,MertensTuriaciVerlinde2017}. It would be interesting to test the coherence between fusion paths
on a quantum processor using local singlet measurements. For example, in the four-spin example, the \(Q\)-singlet probability of the middle pair distinguishes the coherent fusion state from the mixed state obtained by sampling the two fusion paths with equal probability. 
The fusion construction extends to other compact Lie groups and yields
quantum channels with nonnegative branching weights. These extensions
may provide a starting point for constructing higher-dimensional bulk
duals, including AdS space.

\begin{samepage}
\section*{Acknowledgments}
Y.S. is supported by the Special Postdoctoral Researcher Program at RIKEN. We thank T. Takayanagi and M. Watanabe for helpful comments.
The authors used ChatGPT-6.1 sol and Claude Fable 5.1 to assist with analytical derivations, numerical checks, verification of references. 
 
\par\end{samepage}


\newpage

\tableofcontents
\newpage

\appendix

\section{The fusion MPS: conventions and reference state}
\label{app:definitions}
\label{app:comments-definitions}
\label{app:mps-definitions}
\label{app:reference-returns}
\label{supp:entanglement}
\label{app:transfer-resonances}
\label{supp:resonances}

\subsection{Representations and CG conventions}
In the compact real form, the representation with $K=Q^{2J_z}$ and
the ordered coproduct are given by~\cite{Jimbo1985,KlimykSchmudgen1997}
\begin{equation}
 \begin{aligned}
 J_+|j,m\rangle&=\sqrt{[j-m]_Q[j+m+1]_Q}|j,m+1\rangle,\\
 J_-&=J_+^\dagger,\quad K^\dagger=K,\\
 \Delta(K)&=K\otimes K,\\
 \Delta(J_\pm)&=J_\pm\otimes K^{1/2}+K^{-1/2}\otimes J_\pm.
 \end{aligned}
 \label{eq:app-mps-generators}
\end{equation}
The generators satisfy $[J_+,J_-]=(K-K^{-1})/(Q-Q^{-1})$, with a continuous
$Q=1$ limit. We denote the real CG coefficients by
$C^{Jm'}_{jm,\alpha}=C_Q(jm;\tfrac12\alpha| Jm')$
in the expansion
$|J,m'\rangle=\sum_{m,\alpha}C^{Jm'}_{jm,\alpha}|j,m\rangle\otimes|\alpha\rangle$,
with $m'=m+\alpha$. They resolve
$V_j\otimes V_{1/2}=V_{j+1/2}\oplus V_{j-1/2}$, with only the upper
branch at $j=0$. In the notation
$C_{\pm,\alpha}(j,m')=C^{j\pm1/2,m'}_{j,m'-\alpha,\alpha}$,
our phase convention gives
\begin{equation}
 \begin{gathered}
 \begin{pmatrix}
 C_{+,1/2}&C_{-,1/2}\\
 C_{+,-1/2}&C_{-,-1/2}
 \end{pmatrix}
 =\begin{pmatrix}c_j^+(m')&-c_j^-(m')\\
 c_j^-(m')&c_j^+(m')\end{pmatrix},\\
 c_j^+(m')=Q^{(j-m'+1/2)/2}
 \sqrt{\frac{[j+m'+1/2]_Q}{d_j}},\\
 c_j^-(m')=Q^{-(j+m'+1/2)/2}
 \sqrt{\frac{[j-m'+1/2]_Q}{d_j}}.
 \end{gathered}
 \label{eq:app-mps-explicit-cg}
\end{equation}
The matrix applies for $|m'|\le j-\tfrac12$. At either extremal $m'$,
the only allowed upper-branch coefficient is $1$. Forbidden entries
vanish. The entries satisfy $[c_j^+(m')]^2+[c_j^-(m')]^2=1$.
The upper highest-weight coefficient and the second component of the
lower column are positive in this phase convention.

\subsection{Normalization and scalar closure}
The CG coefficients satisfy three orthogonality identities in the
ordered fusion basis,
\begin{equation}
 \begin{aligned}
 \sum_{m,\alpha}C^{Jm'}_{jm,\alpha}C^{Lm''}_{jm,\alpha}
 &=\delta_{JL}\delta_{m'm''},\\
 \sum_{\alpha,m'}Q^{2m'}C^{Jm'}_{jm,\alpha}C^{Jm'}_{jk,\alpha}
 &=\delta_{mk}Q^{2m}\frac{d_J}{d_j},\\
 \sum_{\alpha}Q^{2\alpha}
 C^{J,m+\alpha}_{jm,\alpha}C^{J,m+\alpha}_{j+1,m,\alpha}
 &=0\quad(J=j+\tfrac12).
 \end{aligned}
 \label{eq:app-mps-cg-orthogonality}
\end{equation}
The last identity cancels the cross terms between inequivalent input
representations. Together with the second identity and $\sum_Jp_{jJ}=1$,
it establishes completeness on finite-support states. The isometry
therefore extends to all of $\mathcal H$. For input $G_j$, the first
identity eliminates coherences between inequivalent output
representations and gives the scalar closure
$\widehat{\mathcal E}(G_j)=\sum_Jp_{jJ}G_J$ of Sec.~\ref{sec:fusion-model}.

\subsection{Reference spectral measure and Hamiltonian returns}
For Markov-trace weights, the spectral measure in the height vacuum has
density
\begin{equation}
 \rho_0(E)=\frac{\sqrt{4E_*^2-E^2}}{2\pi E_*^2},
 \quad |E|<2E_*,
\end{equation}
with zero density outside the stated interval. For general allowed
branch weights, $H_{\rm eff}=E_0\mathcal J_a$ is bounded because
$a_n\le\delta$. The Taylor series of the return amplitude therefore
defines an entire function,
\begin{equation}
 \mathcal A(t)=\langle0|e^{-itH_{\rm eff}/\hbar}|0\rangle
 =\sum_{N=0}^{\infty}\frac{(-it/\hbar)^N}{N!}
       (E_0\delta)^N Z_N .
\end{equation}
Here $Z_N$ is the cap success probability and $|\mathcal A(t)|^2$ is
the Hamiltonian return probability. The fusion length $N$ counts
spatial sites, and $t$ denotes Hamiltonian evolution time.
The series uses the cap probabilities at all lengths generated by the
same fusion tensors.

\subsection{Physical-spin entanglement of the Markov reference}
For even $N$ and finite $Q\ge1$, we consider the normalized physical-spin
ket with Markov-trace branch weights and vacuum caps. The Schmidt
decomposition uses the ordinary positive inner product. A cut after
$L$ sites separates blocks with $0<L<N$ and $R=N-L$. The ballot number
\begin{equation}
 D_a(h)=\binom a{(a-h)/2}-\binom a{(a-h)/2-1}
\end{equation}
counts nonnegative $a$-step paths from zero to height $h$ and vanishes
unless $0\le h\le a$ and $h\equiv a\pmod2$. The nonzero eigenvalues of
the reduced density matrix are
\begin{equation}
 \begin{aligned}
 w_j&=\frac{D_L(2j)D_R(2j)}{\mathrm{Cat}_{N/2}},\\
 \lambda_{j,m}&=w_j\frac{Q^{2m}}{d_j},\quad -j\le m\le j.
 \end{aligned}
\label{eq:sm-markov-schmidt}
\end{equation}
The cut spin satisfies $2j\equiv L\pmod2$ and
$0\le2j\le\min(L,R)$.

The equal fusion-path amplitudes and CG orthogonality leave one coherent
multiplicity vector for each $(j,m)$. In an orthonormal suffix basis,
the invariant pair across the cut takes the form
$\sum_m(-1)^{j-m}Q^m|j,m\rangle_L|j,-m\rangle_R/\sqrt{d_j}$.
The counts of compatible prefixes and suffixes give
Eq.~\eqref{eq:sm-markov-schmidt}. The identities $\sum_jw_j=1$ and
$\sum_mQ^{2m}=d_j$ fix its normalization.

The cut-height distribution is independent of $Q$. The dependence on
the deformation enters through the magnetic weights and hence the
physical-spin entanglement. At $Q=1$, the weights within each multiplet
are all $w_j/(2j+1)$. The entropy is the sum of the Shannon entropy of
$w_j$ and the average entropy of $G_j$. Each allowed magnetic label
gives one Schmidt value, and the quantum dimension normalizes the
weights within its multiplet.

\subsection{Operator ranks}
The space $\operatorname{Hom}(V_l,V_j)$ consists of linear maps
$V_l\to V_j$. For the raw CG tensors $\sqrt{p_{jJ}}C_Q$, rank $s$
labels the spin-$s$ summand of $V_j\otimes V_l^*$, where $V_l^*$
is the dual representation. The map $\mathcal D_{jl}(T)=G_j^{1/2}TG_l^{1/2}$
transports this decomposition to the trace-preserving tensors on
finite-support operators. It takes the rank-zero block spanned by
$I_{V_j}$ to the block with trace-normalized representative $G_j$.

\subsection{Transfer resonances beyond the invariant sector}
For Markov-trace weights, $\mathrm{Hom}(V_l,V_j)$ decomposes into ranks
$s=|j-l|,\ldots,j+l$. The transfer preserves $s$ and its magnetic
component. The same matrix $W^{(s)}$ on pairs $(j,l)$ acts for each
magnetic component. The diagonal return from $(j,l)=(0,s)$ and its
resolvent are
\begin{equation}
\begin{gathered}
 C_s(N)=[(W^{(s)})^N]_{(0,s),(0,s)},\\
 R_s(z)=\sum_{N\ge0}\frac{C_s(N)}{z^{N+1}}.
\end{gathered}
\label{eq:proposal-sector-return}
\end{equation}
The series defines a local resolvent at sufficiently large $|z|$.
We continue this matrix element on the infinite tower. The spectral
cut and sheets below belong to the local continuation of the selected
matrix element of the CPTP map.
For $s=0$, $C_0(N)=Z_N$. For $s>0$, the return measures the response of
a coherence mapping $V_s$ to $V_0$.
The Jacobi cut $[-2/\delta,2/\delta]$ of the scalar return is uniformized
by the coordinate $w$,
\begin{equation}
 z=\frac{w+w^{-1}}{\delta},\quad
 \begin{cases}|w|>1&\text{first sheet},\\|w|<1&\text{second sheet}.
 \end{cases}
\label{eq:proposal-resonance-sheets}
\end{equation}
The cut maps to $|w|=1$. The poles on the second sheet define the
transfer resonances of the infinite tower.

At large $Q$, a shell of fixed $j+l$ contains $2s+1$ transverse values
of $l-j$. Its leading recurrence is that of a Dirichlet path graph,
which predicts the first-quadrant representatives
\begin{equation}
\begin{gathered}
 |w_\nu|\sim Q^{-1/2},\quad
 \arg w_\nu\sim\frac{\pi\nu}{2(2s+2)},\\
 1\le\nu\le2s,\quad \nu\equiv2s\pmod2.
\end{gathered}
\label{eq:prl-resonance-leading}
\end{equation}
Sign reversal and complex conjugation generate the remaining poles.
High-precision numerical continuation recovers this pattern for
$s=\tfrac12,1,\tfrac32,2,\tfrac52$ under changes of the continuation
procedure and the closure of the infinite tail. The scalar resolvent
has no finite nonzero-$w$ pole. The prediction uses asymptotic shell
separation, and the numerical checks cover the listed ranks with
Markov-trace weights. A pure $q$-oscillator fitted to the second and
fourth nonscalar returns fails at the sixth. Nonscalar returns can
also fail the positivity conditions for a Jacobi spectral measure.
A gravitational quasinormal interpretation requires a specified bulk
time generator and continuation.

\section{A static SYK interpolation with the same disorder}
\label{supp:syk-interpolation}

Fix $0<q_*<1$ and choose even $p$ with
$2p^2/N_f\to\Lambda_*=-\log q_*$. Take an even
$M_{\min}\asymp N_f^\alpha$ with $3/4<\alpha<1$, and set
$M_{N_f}(u)=(1-u)M_{\min}+uN_f$ for $0\le u\le1$.
When $M=M_{N_f}(u)$ is even, let only the first $M$ Majoranas interact,
\begin{equation}
 H_{N_f}(u)=\frac{E_*}{\sqrt{\binom{M}{p}}}
 \sum_{\substack{I\subseteq\{1,\ldots,M\}\\|I|=p}}J_I\Gamma_I,
 \quad \Gamma_I=i^{p/2}\prod_{i\in I}\chi_i ,
\label{eq:sm-static-syk-family}
\end{equation}
with the product in increasing order. The Hilbert space and Gaussian
couplings $J_I$ are shared for all $u$. Between consecutive even values
of $M$, interpolate the unnormalized sums linearly and rescale to keep
$\mathbb E\operatorname{tr}H_{N_f}(u)^2=E_*^2$. This gives a continuous
family at finite $N_f$.

In each Wick pairing the supports are sampled with their squared
weights. For the interpolated sum, the probability that a support uses
either of the two added Majoranas is at most $2p/(M+2)$, with $M$ the
smaller core size. Conditioning on supports within this core gives the
ordinary SYK moment. Since normalized Clifford traces have magnitude at
most one, the error in a fixed moment is $O_k(p/M_{\min})=o(1)$.

Along any sequence $u_{N_f}\to u>0$, $M/N_f\to u$, so
$2p^2/M\to\Lambda_*/u$ and the crossing weight is
$q(u)=e^{-\Lambda_*/u}$. For $u_{N_f}\to0$, we have $p^2/M\to\infty$
and $p^3/M^2\le p^3/M_{\min}^2\to0$, giving $q(0)=0$.
The DSSYK and noncrossing moment limits then give~\cite{BINT2019,FengTianWei2019}
\begin{equation}
 \frac{\mathbb E\operatorname{tr}H_{N_f}(u)^{2k}}{E_*^{2k}}
 \longrightarrow m_{2k}(q(u))=\frac{Z_{2k}(q(u))}{Z_2(q(u))^k},
\label{eq:sm-static-syk-uniform-moments}
\end{equation}
where $Z_N(q)$ is the MPS return probability for the branch weights in
Eq.~\eqref{eq:prl-dssyk-weights}. The ratio follows by normalizing its
second moment in Eq.~\eqref{eq:proposal-hamiltonian-moments}. The convergence
is uniform in $u$ by compactness and continuity of the limiting
moments. Odd disorder-averaged moments vanish. Thus $u=0$ gives fusion
SYK and $u=1$ gives DSSYK with $q=q_*$.

An earlier semicircle argument treated $p/N_f=1/2$~\cite{LiuNowakZahed2017},
and Ref.~\cite{RazYoussef2026} studied the $q=0$ spectral form factor.

\section{Poisson sigma model and the dilaton-gravity action}
\label{app:psm}
\label{app:gravity-definitions}

This appendix gives the steps from the charge algebra to the
action~\eqref{prl:eq:sinhaction}.

In the notation of Appendix~\ref{app:mps-definitions}, $E=J_+$, $F=J_-$,
and $K=Q^{2J_z}$, so that $[J_z,J_\pm]=\pm J_\pm$ and
$[J_+,J_-]=\sinh(2\gamma J_z)/\sinh\gamma$ with $\gamma=\log Q$.
For $\widehat\phi=h_gJ_z$, $\widehat Y_1=h_g(J_++J_-)/2$, and
$\widehat Y_2=h_g(J_+-J_-)/(2i)$, with $\gamma=\lambda h_g/2$, this gives
\begin{equation}
\begin{gathered}[c]
 [\widehat\phi,\widehat Y_1]=ih_g\widehat Y_2,\quad
 [\widehat\phi,\widehat Y_2]=-ih_g\widehat Y_1,\\
 [\widehat Y_1,\widehat Y_2]
 =ih_g\,\frac{h_g/2}{\sinh(\lambda h_g/2)}\,\sinh(\lambda\widehat\phi).
\end{gathered}
\label{eq:app-psm-commutators}
\end{equation}
The rescaled commutator defines the classical bracket
$\{f,g\}=\lim_{h_g\to0}[\widehat f,\widehat g]/(ih_g)$, with $h_gm$,
$h_g(j+\tfrac12)$, and $\lambda$ fixed. It gives
$\{\phi,Y_1\}=Y_2$, $\{\phi,Y_2\}=-Y_1$, and
$\{Y_1,Y_2\}=V_\lambda(\phi)=\sinh(\lambda\phi)/\lambda$.

For target fields $X^i=(\phi,Y_1,Y_2)$, the bracket takes the form
$\{f,g\}=P^{ij}(X)\partial_i f\,\partial_jg$, with $P^{\phi1}=Y_2$,
$P^{\phi2}=-Y_1$, and $P^{12}=V_\lambda(\phi)$.
With spacetime one-forms $A_i$, the Poisson sigma model has
action~\cite{SchallerStrobl1994,Ikeda1994}
\begin{equation}
 I_{\rm PSM}=A_g\int_\Sigma
 \left[A_i\wedge dX^i+\frac12P^{ij}A_i\wedge A_j\right].
\label{eq:gravity-psm}
\end{equation}
The gravitational identification is $A_\phi=\omega$ and $A_{Y_a}=e^a$.
Up to boundary terms, the integrand then becomes
\begin{equation}
 \phi\,d\omega+V_\lambda(\phi)\,e^1\wedge e^2+Y_1T^1+Y_2T^2,
\label{eq:app-psm-first-order}
\end{equation}
with $T^1=de^1-\omega\wedge e^2$ and $T^2=de^2+\omega\wedge e^1$.
The $Y_a$ equations impose $T^1=T^2=0$.
For a nondegenerate frame, $g_{\mu\nu}=e^a_\mu e^a_\nu$ and
$d\omega=-Re^1\wedge e^2/2$ give the metric action of
Eq.~\eqref{prl:eq:sinhaction}.
The boundary action $I_\partial$ fixes the induced metric and dilaton
and includes the counterterms.

The equation of motion of $\phi$ is $R=2V_\lambda'(\phi)=2\cosh(\lambda\phi)$.
The curvature is therefore constant, $R=2$, at vanishing dilaton for any
$\lambda$, and for any dilaton at $\lambda=0$.

The bulk action follows from the bracket. The inputs beyond the bracket
are the scaling limit, the assignment of $\phi$ to the dilaton and of
$A_{Y_a}=e^a$ to the frame, the normalization $A_g$, and the boundary term.
Perturbative quantization of the Poisson sigma model gives a deformation
quantization of the bracket~\cite{CattaneoFelder2000}.

\section{dSJT amplitudes in the CJ prescription}
\label{app:disk-seeds}
\label{app:seeds}

\begin{figure}[!tbp]
\centering
\begingroup
\definecolor{gravityInk}{RGB}{36,62,82}
\definecolor{gravityTeal}{RGB}{40,125,142}
\definecolor{gravityLight}{RGB}{238,243,246}
\definecolor{gravityMuted}{RGB}{111,122,131}
\begin{tikzpicture}[x=1cm,y=1cm,scale=.90,
 every node/.style={font=\fontsize{8}{9.2}\selectfont,text=gravityInk},
 draw=gravityInk,line width=.65pt,line cap=round,line join=round]
 \path[use as bounding box] (0,-1.08) rectangle (7.9,3.35);

 \path[fill=gravityLight,draw=gravityInk]
   (3.95,.60) ellipse (3.65 and .82);
 \foreach \x/\i in {1.30/1,3.80/2,6.55/n_{\rm b}}{
  \pgfmathsetmacro{\seam}{.60+.82*sqrt(1-((\x-3.95)/3.65)^2)}
  \path[fill=white,draw=none]
   (\x-.30,\seam)
   .. controls (\x-.34,\seam+.48) and (\x-.78,2.32) .. (\x-.95,2.75)
   -- (\x+.95,2.75)
   .. controls (\x+.78,2.32) and (\x+.34,\seam+.48) .. (\x+.30,\seam)
   -- cycle;
  \draw
   (\x-.30,\seam)
   .. controls (\x-.34,\seam+.48) and (\x-.78,2.32) .. (\x-.95,2.75);
  \draw
   (\x+.30,\seam)
   .. controls (\x+.34,\seam+.48) and (\x+.78,2.32) .. (\x+.95,2.75);
  \draw[fill=white] (\x,2.75) ellipse (.95 and .13);
  \draw[fill=white,draw=gravityTeal,line width=.85pt]
   (\x,\seam) ellipse (.30 and .08);
  \node at (\x,3.16) {$\tau_{\i}$};
  \node at (\x,2.43) {$T_{\tau_{\i}}(P_{\i})$};
  \node[text=gravityTeal] at (\x,\seam-.24) {$P_{\i}$};
 }
 \node[text=gravityMuted] at (5.18,2.43) {$\cdots$};

 \begin{scope}[yshift=.10cm]
 \path[fill=gravityLight,draw=gravityInk]
  (5.43,.43)
  .. controls (5.65,.88) and (6.33,.88) .. (6.55,.43)
  .. controls (6.36,.23) and (5.62,.23) .. cycle;
 \draw[fill=white] (5.99,.49) ellipse (.31 and .10);
 \end{scope}
 \node at (5.75,.16) {genus $g$};
 \node[font=\fontsize{9}{10}\selectfont] at (3.15,.51)
  {$K_{g,n_{\rm b}}(\boldsymbol P)$};
 \node[text=gravityMuted] at (3.15,.04) {$n_{\rm b}$ boundaries};

 \node at (3.95,-.78)
  {\textcolor{gravityMuted}{Euler weight}\quad
   $g_s^{\,2g-2+n_{\rm b}}\prod_{i=1}^{n_{\rm b}}dP_i
   \quad(P_i>0)$};
\end{tikzpicture}
\endgroup
\caption{The connected dSJT amplitude is obtained by sewing
$n_{\rm b}$ boundary trumpets $T_{\tau_i}$ to a genus-$g$ core $K_{g,n_{\rm b}}$
and integrating each squared neck size $P_i>0$.
The Euler weight is $g_s^{2g-2+n_{\rm b}}$. Adding a handle contributes $g_s^2$.
The core includes the CJ moduli measure $\operatorname{Pf}(-\Omega)$.}
\label{fig:gravity-sewing}
\end{figure}
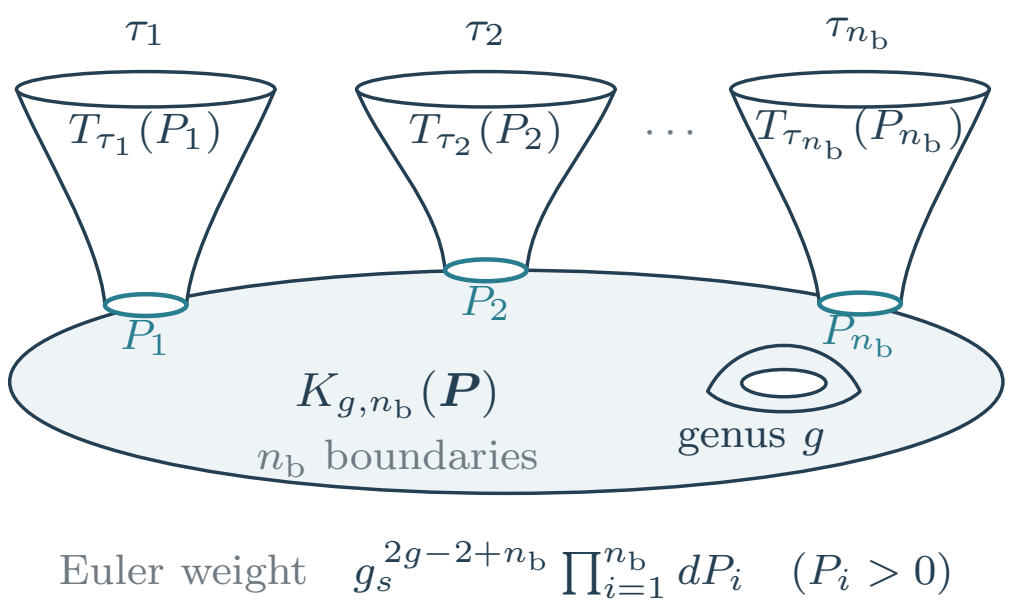

The problem of the path integral in dSJT is that there are no real
Lorentzian solutions for higher genus geometries if we mimic the AdSJT
case~\cite{SaadShenkerStanford2019}. In the Cotler--Jensen (CJ)
prescription~\cite{CotlerJensen2024}, one uses the $(-,-)$ signature and
realizes $R=2$.
On the future boundary, we write $\Phi_F=4\pi^2\varphi_b/\ell_b$, where
$\ell_b$ is the regulated proper boundary length. The complex boundary
parameters are $\tau_{F/P}=(0^+\mp i\Phi_{F/P})^{-1}$ for future and past
circles. The conditions
$\operatorname{Im}\Phi_F>0$ and $\operatorname{Im}\Phi_P<0$ ensure
convergence. The difference from Ref.~\cite{SaadShenkerStanford2019} is that
the measure for the Weil--Petersson symplectic form $\Omega$ is given by
$\operatorname{Pf}(-\Omega)=(-1)^{d_{\rm mod}}\operatorname{Pf}(\Omega)$,
where $d_{\rm mod}=3g-3+n_{\rm b}$ is half the real moduli-space dimension,
and that the geodesic length is continued to
$b=2\pi i\sqrt P$ with $P\in[0,\infty)$. At genus $g$ with $n_{\rm b}$
boundaries, the connected amplitude is
\begin{equation}
\begin{aligned}
 \mathcal Z_{g,n_{\rm b}}
 &=g_s^{2g-2+n_{\rm b}}\int_{\mathbb R_+^{n_{\rm b}}}
       \prod_i[dP_i\,T_{\tau_i}(P_i)]\,K_{g,n_{\rm b}},\\
 K_{g,n_{\rm b}}&=(-\alpha_{\rm JT})^{d_{\rm mod}}V_{g,n_{\rm b}}(2\pi i\sqrt{\boldsymbol P}),\quad
 T_\tau(P)=\frac{e^{-P/\tau}}{\sqrt{2\pi\tau}}.
\end{aligned}
\label{eq:gravity-all-topology}
\end{equation}
Here $g_s=e^{-S_0}$, $d_{\rm mod}=3g-3+n_{\rm b}$, $\alpha_{\rm JT}=(2\pi^2)^{-1}$,
$2g-2+n_{\rm b}>0$, and $V_{g,n_{\rm b}}$ is the standard Weil--Petersson volume~\cite{Mirzakhani2007}.
Figure~\ref{fig:gravity-sewing} illustrates the boundary sewing.

The remaining cases, $2g-2+n_{\rm b}\le0$, are the disk and the cylinder.
The disk seed is \(g_s^{-1}\delta'(P-1)\), and the cylinder seed is
\(\delta(P_1-P_2)\). Their pairings with the regulated trumpet yield
the amplitudes
\begin{equation}
\begin{gathered}
 Z_{0,1}=\frac{e^{-1/\tau}}{\sqrt{2\pi}g_s\tau^{3/2}},\quad
 Z_{0,2}=\frac{\sqrt{\tau_1}\sqrt{\tau_2}}
                   {2\pi(\tau_1+\tau_2)} .
\end{gathered}
\label{eq:endmatter-disk-trumpet}
\end{equation}
The disk pairing evaluates to \(-g_s^{-1}T_\tau'(1)\).
The cylinder pairing integrates \(T_{\tau_1}T_{\tau_2}\) over \(P>0\).
In the main text all boundaries are future circles, for which
\(\tau=\tau_F=(0^+-i\Phi_F)^{-1}\).

\section{Collective preparation and genus expansion}
\label{supp:matrix-preparation}
\label{supp:same-disorder}

This appendix shows how the matrix integral of the main text arises as a
collective measurement on $M$ copies, first in the height space and then
in SYK with shared disorder, and then gives its planar limit and genus
expansion. What is prepared is the form of the measure,
$\Delta(\boldsymbol x)^2\prod_ie^{-V_\epsilon(x_i)/\hbar_{\rm m}}$.
The potential $V_\epsilon$, the number of copies $M$, and the
antisymmetrization are inputs; they are not derived from the fusion MPS.

\subsection{Preparation in the height space}
Here $M$ counts independent copies and $\mathfrak t=M\hbar_{\rm m}$
denotes the matrix coupling. The fusion length and physical time are
separate variables. The scalar return is
$\mathcal R(z)=\sum_{k\ge0}z^{2k}/[k!(k+1)!]$.
The state
$|\phi\rangle=\sqrt{\pi/2}(4-W^2)^{-1/4}|0\rangle$
has the flat spectral measure $dw/4$. We set $x=s_\epsilon w$. The Legendre orbitals
$\sqrt{2j+1}P_j(W/2)|\phi\rangle$, where $P_j$ is a Legendre polynomial,
are orthonormal with respect to this measure. Their $M$-copy Slater
determinant is proportional to the Vandermonde
$\Delta(\boldsymbol x)=\prod_{i<j}(x_i-x_j)$.
The common weight and multiplication source $\eta$ give the normalized
functional
\begin{equation}
 Z_{\epsilon,M}[\eta]=
 \frac{\int_{\mathcal C^M}\Delta(\boldsymbol x)^2
       \prod_i e^{-V_\epsilon(x_i)/\hbar_{\rm m}+\eta(x_i)}dx_i}
      {\int_{\mathcal C^M}\Delta(\boldsymbol x)^2
       \prod_i e^{-V_\epsilon(x_i)/\hbar_{\rm m}}dx_i}.
\label{eq:gravity-matrix-preparation}
\end{equation}
The real contour is $\mathcal C=\mathcal C_\epsilon=[-2s_\epsilon,2s_\epsilon]$,
where $s_\epsilon^2=1+b_{\rm wall}\epsilon^2$ and $0<b_{\rm wall}<1/4$. A positive weight
defines an ordinary collective expectation in the original Hilbert
space. Complex contours specify the analytic continuation of the
moments, and the original inner product is positive.

\subsection{Preparation in SYK with shared disorder}
The collective preparation uses $M$ operator copies of a microscopic
SYK sample $h_J=H_{\rm SYK}(J)/E_*$. Each copy carries the
Hilbert--Schmidt inner product $\operatorname{tr}(X^\dagger Y)$.
The copies share the Gaussian coupling array $J$, so the joint inner
product averages the product of traces over $J$.

We hold $M$, $\epsilon>0$, $\hbar_{\rm m}>0$, and the real walls fixed.
For $\nu>0$ and $w=\min(2,\max(-2,\lambda))$, the bounded orbitals are
\begin{equation}
\begin{gathered}
f_\nu(w)=\sqrt{\pi/2}\,(4-w^2+\nu)^{-1/4},\\
a_{j,\nu}(\lambda)=e^{-V_\epsilon(s_\epsilon w)/(2\hbar_{\rm m})}
 \sqrt{2j+1}\,P_j(w/2)f_\nu(w),
\end{gathered}
\label{eq:same-disorder-orbitals}
\end{equation}
where $0\le j<M$ and 
$s_\epsilon$ is fixed by the real walls above.
We apply the clipping inside the orbital functions and keep the
original sample as the spectral operator. The Gram matrix
of $v_j(J)=a_{j,\nu}(h_J)$ determines the regularized orbitals and the
Slater vector,
\begin{equation}
\begin{gathered}
(\mathsf C_J)_{ab}=\operatorname{tr}(v_a^\dagger v_b),\\
q_j=\sum_a v_a[(\mathsf C_J+\mu I)^{-1/2}]_{aj},\\
\Psi_J=q_0\wedge\cdots\wedge q_{M-1},\quad \mu>0 .
\end{gathered}
\label{eq:same-disorder-gram}
\end{equation}
The wedge antisymmetrizes the orbitals with the factor $1/\sqrt{M!}$.
For the spectral observable $O_J=F(h_J^{(1)},\ldots,h_J^{(M)})$ with
bounded continuous $F$, the Slater norm and prepared expectation are
\begin{equation}
\begin{gathered}
\|\Psi_J\|^2=\frac{\det\mathsf C_J}
 {\det(\mathsf C_J+\mu I)}\le1,\\
\mathcal R_{\nu,\mu}(F)=
\frac{\mathbb E_J\langle\Psi_J,O_J\Psi_J\rangle}
 {\mathbb E_J\|\Psi_J\|^2}.
\end{gathered}
\label{eq:same-disorder-readout}
\end{equation}
We average the numerator and norm separately before taking their
ratio. At fixed $M$, the denominator is positive for sufficiently
large microscopic size.

For a fixed Lipschitz function $f$, the Gaussian Poincar\'e inequality
and Clifford orthogonality bound the sample variance,
\begin{equation}
\operatorname{Var}_J[\operatorname{tr}f(h_J)]
\le\frac{\mathbb E_J\operatorname{tr}|f'(h_J)|^2}{\binom{N_f}{p}} .
\label{eq:same-disorder-concentration}
\end{equation}
Together with the microscopic semicircle limit in
Eq.~\eqref{eq:proposal-markov-microscopic-limit}, this bound implies
convergence of the finite Gram matrix to a positive deterministic limit.
We take the microscopic limit at fixed regulators and then let
$\mu\downarrow0$ and $\nu\downarrow0$.
The resulting Slater spectral measure is
\begin{equation}
\begin{gathered}
Z_M^{-1}\Delta(\boldsymbol\lambda)^2
\prod_i e^{-V_\epsilon(s_\epsilon\lambda_i)/\hbar_{\rm m}}d\lambda_i,
\\ \lambda_i\in[-2,2].
\end{gathered}
\label{eq:same-disorder-prepared-law}
\end{equation}
Under $x_i=s_\epsilon\lambda_i$, this measure coincides with the real
preparation in Eq.~\eqref{eq:gravity-matrix-preparation}. It describes
the $M$ energies measured in the prepared state. Microscopic convergence
holds at fixed $M$ and fixed matrix regulators. A simultaneous large-$M$
or matrix critical limit would require uniform estimates.

\subsection{Planar limit}
The planar density $\rho_{\mathfrak t}$ is normalized by
$\int\rho_{\mathfrak t}=\mathfrak t$. For a symmetric one-cut support
$[-2\sqrt r,2\sqrt r]$, the planar equilibrium
equations~\cite{BessisItzyksonZuber1980,BalasubramanianMaganWu2023} give
\begin{equation}
 V_\epsilon'(x)=2\,\mathrm{PV}\int
 \frac{\rho_{\mathfrak t}(y)}{x-y}dy,\quad
 \partial_{\mathfrak t}\rho_{\mathfrak t}(x)=
 \frac1{\pi\sqrt{4r-x^2}}.
\label{eq:gravity-equilibrium}
\end{equation}
Here PV denotes the principal value. Symmetrizing the force-balance
equation gives
$\int xV_\epsilon'\rho_{\mathfrak t}=\mathfrak t^2$.
Differentiation and the arcsine moments give the potential coefficients
in the main text. The chosen return prescription yields
$\mathfrak t'(r)=I_0(2\pi\sqrt{1-r}/\epsilon)/2$, which gives the
critical planar density
\begin{equation}
 \rho_{\mathfrak t_c}(x)=\int_{x^2/4}^1
 \frac{\mathfrak t'(r)dr}{\pi\sqrt{4r-x^2}}
 =\frac{\epsilon}{4\pi^2}
 \sinh\!\left(\frac\pi\epsilon\sqrt{4-x^2}\right),
\label{eq:gravity-planar-density}
\end{equation}
for $|x|\le2$. Division by $\hbar_{\rm m}$ and inclusion of the edge
Jacobian give the planar number density in the main text. This density
is the disk coefficient of the genus expansion.

\section{Boundary reparametrizations from SYK and the MPS}
\label{supp:boundary-reparam}
\label{app:reparametrization}

This appendix gives the details of the Schwarzian paragraph of
Sec.~\ref{sec:gravity-2d}. Throughout, $h_f=H_{\rm SYK}/E_*$ and
\begin{equation}
    S|n\rangle=|n+1\rangle,\quad W=S+S^\dagger,\quad\hat n|n\rangle=n|n\rangle,\quad R=\hat n+1.
\end{equation}
In this appendix $a$ and $b$ denote heights, and $\lambda$ and $E$ denote
Laplace variables; they are unrelated to the symbols of the main text.

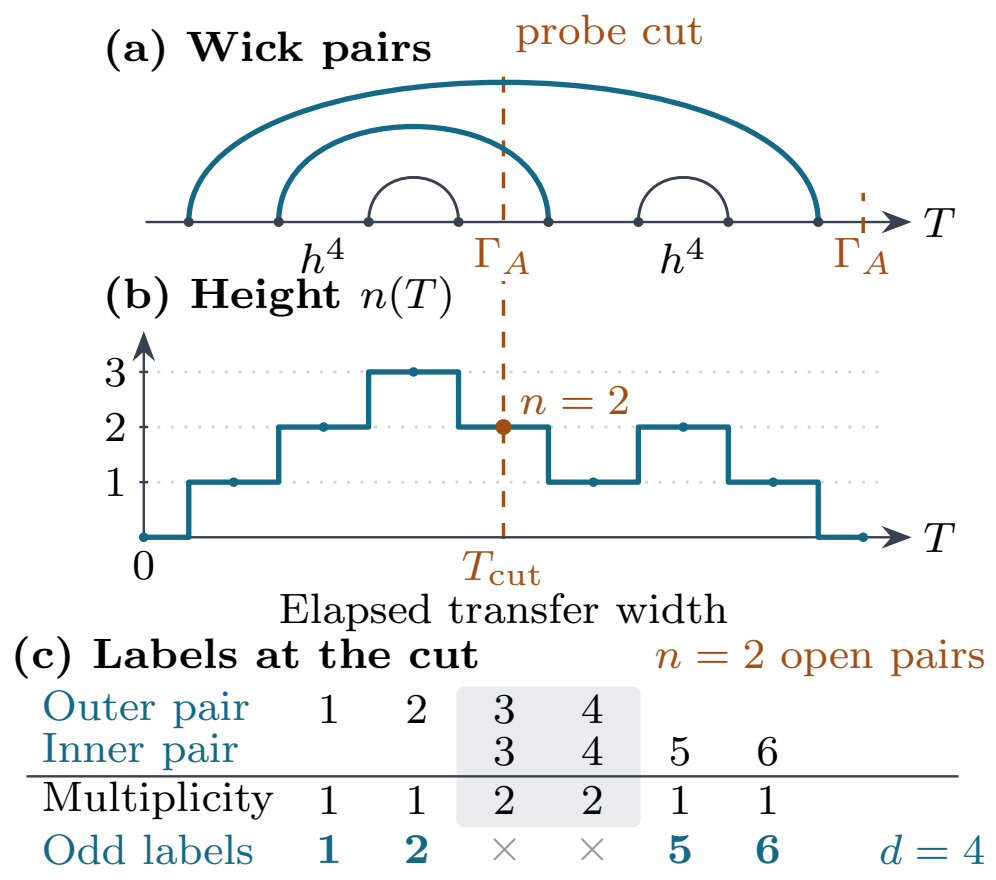
\begin{figure}[!t]
\centering
\begingroup
\definecolor{WickBlue}{HTML}{146986}
\definecolor{WickCut}{HTML}{A24F16}
\definecolor{WickInk}{HTML}{374151}
\definecolor{WickGrid}{HTML}{CED5DC}
\begin{tikzpicture}[
  x=.78cm,y=.49cm,scale=.93,font=\footnotesize,
  >={Stealth[length=2.0mm]},
  axis/.style={draw=WickInk,line width=.45pt},
  open pair/.style={draw=WickBlue,line width=1pt},
  closed pair/.style={draw=WickInk,line width=.55pt},
  cut/.style={draw=WickCut,dashed,line width=.65pt},
  height path/.style={draw=WickBlue,line width=1pt,line join=round}
]
  \begin{scope}[shift={(.81,0)},xscale=.82,yscale=.80]
  \draw[cut] (4.5,-.03)--(4.5,8.46);

  \node[anchor=west,font=\footnotesize\bfseries] at (-.15,8.83)
    {(a) Wick pairs};
  \node[anchor=south west,inner xsep=2pt,text=WickCut] at (4.5,8.49)
    {probe cut};
  \draw[axis,->] (.52,5.72)--(9.04,5.72)
    node[right,inner sep=2pt] {$T$};
  \draw[open pair] (1,5.72) .. controls (1,9.10) and (8,9.10) .. (8,5.72);
  \draw[open pair] (2,5.72) .. controls (2,8.03) and (5,8.03) .. (5,5.72);
  \draw[closed pair] (3,5.72) .. controls (3,6.80) and (4,6.80) .. (4,5.72);
  \draw[closed pair] (6,5.72) .. controls (6,6.80) and (7,6.80) .. (7,5.72);
  \foreach \i in {1,...,8}
    \fill[WickInk] (\i,5.72) circle (1.25pt);
  \draw[cut] (8.5,5.52)--(8.5,6.25);
  \node[below,inner sep=1.2pt] at (2.5,5.57) {$h^4$};
  \node[below,inner sep=1.2pt,text=WickCut,fill=white] at (4.5,5.56)
    {$\Gamma_A$};
  \node[below,inner sep=1.2pt] at (6.5,5.57) {$h^4$};
  \node[below,inner sep=1.2pt,text=WickCut] at (8.5,5.56) {$\Gamma_A$};

  \node[anchor=west,font=\footnotesize\bfseries] at (-.15,4.35)
    {(b) Height $n(T)$};
  \foreach \n in {1,2,3} {
    \draw[WickGrid,dotted,line width=.55pt] (.5,\n)--(8.69,\n);
    \draw[axis] (.45,\n)--(.55,\n);
    \node[anchor=east,inner sep=1.8pt] at (.43,\n) {$\n$};
  }
  \draw[axis,->] (.5,0)--(.5,3.74);
  \draw[axis,->] (.5,0)--(9.04,0)
    node[right,inner sep=2pt] {$T$};
  \draw[height path] (.5,0)--(1,0)--(1,1)--(2,1)--(2,2)
    --(3,2)--(3,3)--(4,3)--(4,2)--(5,2)--(5,1)
    --(6,1)--(6,2)--(7,2)--(7,1)--(8,1)--(8,0)--(8.5,0);
  \foreach \j/\n in {0/0,1/1,2/2,3/3,4/2,5/1,6/2,7/1,8/0}
    \fill[WickBlue] ({\j+.5},\n) circle (1.2pt);
  \node[below,inner sep=1.5pt] at (.5,-.09) {$0$};
  \node[below,text=WickCut,inner sep=1.5pt] at (4.5,-.09) {$T_{\rm cut}$};
  \fill[WickCut] (4.5,2) circle (2pt);
  \node[above right,text=WickCut,inner sep=2pt] at (4.55,2.02) {$n=2$};
  \node[below,inner sep=1pt] at (4.5,-.90) {Elapsed transfer width};
  \end{scope}

  \node[anchor=west,font=\footnotesize\bfseries] at (-.15,-1.75)
    {(c) Labels at the cut};
  \node[anchor=east,text=WickCut] at (9.05,-1.75) {$n=2$ open pairs};
  \fill[WickGrid!45,rounded corners=1.5pt] (4.07,-2.17) rectangle (5.75,-4.21);
  \node[anchor=west,text=WickBlue] at (.13,-2.50) {Outer pair};
  \node[anchor=west,text=WickBlue] at (.13,-3.11) {Inner pair};
  \node[anchor=west] at (.13,-3.82) {Multiplicity};
  \node[anchor=west,text=WickBlue] at (.13,-4.53) {Odd labels};
  \draw[axis] (.15,-3.47)--(8.65,-3.47);
  \foreach \i in {1,2,3,4}
    \node at ({2.91+.8*(\i-1)},-2.50) {$\i$};
  \foreach \i in {3,4,5,6}
    \node at ({2.91+.8*(\i-1)},-3.11) {$\i$};
  \foreach \i/\m in {1/1,2/1,3/2,4/2,5/1,6/1}
    \node at ({2.91+.8*(\i-1)},-3.82) {$\m$};
  \foreach \i in {1,2,5,6}
    \node[text=WickBlue] at ({2.91+.8*(\i-1)},-4.53) {$\mathbf{\i}$};
  \foreach \i in {3,4}
    \node[text=WickInk!55] at ({2.91+.8*(\i-1)},-4.53) {$\times$};
  \node[anchor=east,text=WickBlue] at (9.05,-4.53) {$d=4$};
\end{tikzpicture}
\endgroup
\caption{(a,b) In $\operatorname{tr}(h^4\Gamma_A h^4\Gamma_A)$,
the first endpoint of each Wick pair raises the height by one and
the second lowers it. (c) At the first $\Gamma_A$, the outer and inner
blue pairs give $n=2$. Take their interaction supports to be
$\{1,2,3,4\}$ and $\{3,4,5,6\}$, respectively. Labels $3,4$ occur
twice and their signs cancel. Labels $1,2,5,6$ occur once, giving
$d=4$. The example probe $A=\{5,9\}$ contains one of these labels
and has sign $-1$. The mean sign of one draw is $1-2d/N_f$.
For a uniform $k$-subset $A$, the sign average is
$(1-2d/N_f)^k+O(k^2/N_f)$. With $p$ Majoranas
per interaction, $d/p\to n$ at fixed $n$ in the noncrossing limit, and
$2pk/N_f\to w$ gives $e^{-wn}$.}
\label{fig:majorana-cut}
\end{figure}

\subsection{Open fermion amplitudes}
The Schwarzian describes reparametrizations in nearly $\mathrm{AdS}_2$
gravity and low-energy SYK~\cite{MaldacenaStanfordYang2016,Jensen2016,MaldacenaStanford2016,BerkoozFrumkinMamroudSeitz2025}.
In fusion SYK, let us consider
$\operatorname{tr}(h_f^{k_1}\Gamma_Ah_f^{k_2}\Gamma_A)$, where
$\Gamma_A=i^{k/2}\prod_{i\in A}\chi_i$ is ordered by increasing index
and $k=|A|$ is even.
Moving $\Gamma_A$ through an interaction string with support $I$, $|I|=p$,
yields the sign $(-1)^{|A\cap I|}$~\cite{GarciaGarciaJiaVerbaarschot2018}.
The signs cancel for a Wick pair closed on one side of the cut, the
position of $\Gamma_A$.
The point is that the number of pairs open at the cut can be regarded as
the height $n=2j$ of the MPS.
Let $d$ count the distinct Majorana indices that occur an odd number
of times among the $n$ strings across the cut.
Indices occurring an even number of times cancel from the exchange sign.
In the noncrossing limit~\eqref{eq:proposal-markov-microscopic-limit}, we find
$d=pn+o(p)$. A uniform average over probe supports of size $k$ with
$2pk/N_f\to w$ then gives
\begin{equation}
 \begin{aligned}
 &\lim\mathbb{E}_{J,A}\operatorname{tr}
    (h_f^{k_1}\Gamma_A h_f^{k_2}\Gamma_A)\\
 &\qquad=\langle0|
 W^{k_1}D(w)W^{k_2}|0\rangle,
 \quad D(w)=e^{-w\hat n},
 \end{aligned}
\label{eq:app-open-amplitude}
\end{equation}
where the average is taken over the couplings $J_I$ and the probe
supports $A$. For details, see Fig.~\ref{fig:majorana-cut}.
Gaussian Wick bounds extend the result to the time expansions at fixed
chain intervals and finitely many cuts. The diagonal tags $D(w)$ therefore
resolve the number of open pairs in the fermion correlators.
The Gaussian probe $O$ of Eq.~\eqref{eq:proposal-paired-probe} inserts the
same operator with $w=2pp_{\rm pr}/N_f$. There the probe order was tied to
$Q$, so that $w=2\log Q$; here $w$ is varied through the probe order.

\subsection{Chain coordinate and continuum limit}
With $D_Q|n\rangle=d_{n/2}|n\rangle$, the Markov transfer of the
MPS is $\mathsf P=\delta^{-1}D_Q W D_Q^{-1}$, a Doob transform of the scalar walk~\cite{CoghiGarrahan2025}. The quantum
dimensions telescope along each path and cancel between vacuum caps.
The diagonal tags commute with $D_Q$, so the readout acts on one MPS
copy. Factoring out the known powers of $\delta$ and summing the numbers
of steps $N$ with the weights $e^{-2T}T^N/N!$ gives
\begin{equation}
 K(T)=\sum_{N\ge0}e^{-2T}\frac{T^N}{N!}W^N=e^{-T(2-W)} .
\label{eq:app-K}
\end{equation}
Since $W^N$ grows as $2^N$, the number of steps is Poisson distributed
with mean $2T$, and $T=N/2$ for long paths.
At the lower SYK edge, height parity sends
$2+W$ to $2-W$ and preserves the tags.

In the continuum limit $\rho=(n+1)/\Lambda$ and $t=T/\Lambda^2$ with
$\Lambda\to\infty$, the operator $T(2-W)$ becomes $-t\,\partial_\rho^2$
with the regular branch at $\rho=0$, and $K$ becomes the kernel of
Brownian motion. A height path $\rho(t)$ then carries the weight
$e^{-\mathcal S}$ with $\mathcal S=\frac14\int dt\,(\partial_t\rho)^2$,
which is Eq.~\eqref{eq:S0} in the rescaled variables.
Under the transformation~\eqref{eq:mobius},
\begin{equation}
 (\partial_Tn)^2\,dT\;\to\;(\partial_Tn)^2\,dT
 -d\!\left[\frac{c\,n^2}{cT+d}\right],
\label{eq:app-mobius-total-derivative}
\end{equation}
so $\mathcal S$ changes only by endpoint
terms~\cite{ChenFradkinWitczakKrempa2017}. Since
$dT\to dT/(cT+d)^2$ and $n^2\to n^2/(cT+d)^2$, the combination
$d\sigma=dT/n^2$ of Eq.~\eqref{eq:invariant-time} is invariant.

\subsection{Invariant time}
To construct a clock whose rate depends on height, we need the joint
amplitudes for heights at successive cuts of the same height path.
We probe these heights by inserting nested Majorana probe pairs
in the SYK trace and averaging their supports independently.
For an open interval with initial and final heights $a$ and $b$
and successive chain intervals $T_1,T_2,T_3$, the resulting amplitude is
$\langle b|K(T_3)D(w_2)K(T_2)D(w_1)K(T_1)|a\rangle$.
Using the identity $R^{-2}=\int_0^\infty we^{-w}D(w)\,dw$, we can construct
the clock $d\sigma=dT/(n+1)^2$, which is the invariant time of the main
text, and its length along a path,
$I=\int_0^T dT'/(n(T')+1)^2$. To resolve the accumulated clock length $I$,
we introduce its Laplace variable $\lambda$ and weight each height path by
$e^{-\lambda I}$~\cite{Kac1949}. Correspondingly,
$1-\lambda\,dT\,R^{-2}$ is inserted at each step. The continuum limit
gives $H_\lambda=-\partial_\rho^2+\lambda/\rho^2$ with the regular
branch at $\rho=0$ and $d\sigma=dt/\rho^2$.

The height probes the clock rate along a height path, and
the inverse Laplace transform in $\lambda$ selects its total
clock length $I$. We now ask how height paths with the same $I$
are weighted. With $u=\sigma/I$, the inverse clock $\zeta(u)=t(u)/I$
gives the chain coordinate as a function of the clock
reading; it is the function $\zeta$ of the main text up to a constant
factor. Since $dt=\rho^2d\sigma$, it satisfies $\zeta'=e^{2x}$ with
$x=\log\rho$~\cite{Lamperti1972}. The weight of the log-height path thus
determines the action for $\zeta$.

\subsection{Sum over chain lengths}
At fixed $I$, the chain length is
$L=T/\Lambda^2=\int_0^I \rho^2\,d\sigma$. We sum over $L$ to obtain an
amplitude specified by $I$ and the endpoint heights.
We regulate the integral over $L$ with $e^{-EL}$, where $E>0$, and write
\begin{equation}
 G_{\lambda,E}(\rho,\rho_a)
 =\int_0^\infty dL\,e^{-EL}
   \langle\rho|e^{-LH_\lambda}|\rho_a\rangle,
\end{equation}
satisfying
$(H_\lambda+E)G_{\lambda,E}=\delta(\rho-\rho_a)$, where the
source comes from the initial kernel at $L=0$.
Writing $G_{\lambda,E}=\sqrt{\rho}\,\psi(x)$ gives the equation
\begin{equation}
 (\mathcal H_E+\lambda)\psi(x)
 =\sqrt{\rho_a}\,\delta(x-x_a),
 \quad
 \mathcal H_E=-\partial_x^2+Ee^{2x}+\tfrac14 .
\end{equation}
The exponential potential in $\mathcal H_E$ has the Liouville form encountered in the effective quantum mechanics of low-energy SYK~\cite{BagretsAltlandKamenev2016}. Liouville quantum mechanics also appears in the correspondence between
the stationary measure of open KPZ and Euclidean JT evolution between
two end-of-the-world branes~\cite{Watanabe:2025rwp}.
Removing the regulator, $E\to0$, factoring out the endpoint factors
$\sqrt\rho$ and $\sqrt{\rho_a}$, and inverting the Laplace transform in
$\lambda$ gives the free heat kernel
\begin{equation}
 A_0(\ell,I)
 =\langle x_b|e^{-I(-\partial_x^2+1/4)}|x_a\rangle
 =\frac{e^{-I/4-\ell^2/(4I)}}{\sqrt{4\pi I}},
\label{eq:app-free-kernel}
\end{equation}
where $\ell=x_b-x_a$. This is Eq.~\eqref{eq:free-propagator}.
Equivalently, the time-changed generator $\rho^2\partial_\rho^2$ becomes
$\partial_x^2-\partial_x$ in the coordinate $x=\log\rho$, and removing the
drift produces the endpoint factors and the constant $\tfrac14$, which
gives the factor $e^{-I/4}$.

\subsection{Schwarzian form}
To read off the weight of a height path, we compose $A_0$ over short
clock intervals and keep the intermediate heights explicit.
For a slice of length $\Delta\sigma_j$, the Gaussian exponent
contributes $(\Delta x_j)^2/(4\Delta\sigma_j)$. The factors
$e^{-\Delta\sigma_j/4}$ combine into $e^{-I/4}$, which is common
to all height paths at fixed $I$. Summing the path-dependent terms
and taking the smooth-path limit gives
\begin{align}
 \mathcal S&=\frac14\int_0^I d\sigma\,(\partial_\sigma x)^2
     =\frac1{4I}\int_0^1du\,x'(u)^2
 \quad(u=\sigma/I)\nonumber\\
 &=-\frac{1}{8I}\int_0^1 du\,\{\zeta,u\}
      +\frac{x'(1)-x'(0)}{4I},\label{eq:app-Sx-schwarzian}
\end{align}
where $\zeta'=e^{2x}$ and
$\{\zeta,u\}=\zeta'''/\zeta'-\tfrac32(\zeta''/\zeta')^2$~\cite{BelokurovShavgulidze2019,BauerschmidtLosevWildemann2024}.
The second line uses $x'=\zeta''/(2\zeta')$, so that
$\{\zeta,u\}=2x''-2x'^2$. This is Eq.~\eqref{eq:Sx-schwarzian}.

\subsection{Circle and disk}
Let $f$ be a smooth lift of an orientation-preserving circle map,
with $f(u+1)=f(u)+1$ and $f'(u)>0$. We represent the circle by
$\zeta_f(u)=C_0+C_1e^{2\pi i f(u)}$, choosing $C_1$ so that
$\zeta_f'(0)=e^{2x(0)}$.
Then $\zeta_f'=e^{2x_f}$ gives
\begin{equation}
 x_f(u)=x(0)+\frac12\log\frac{f'(u)}{f'(0)}
             +i\pi[f(u)-f(0)] .
\end{equation}
After one circuit, $f'$ returns to its initial value and $f$
increases by one. The resulting endpoint displacement is therefore
$x_f(1)-x_f(0)=i\pi$.

In the open kernel, we keep the endpoint displacement $\ell$
variable and write $P=-\ell^2/\pi^2$. Real endpoints give $P\leq0$,
and the circle corresponds to $P=1$ on the branch $\ell=i\pi$.
In the momentum representation,
$A_0=e^{-I/4}\int_0^\infty e^{-Ik^2}\cos(k\ell)\,dk/\pi$.
The disk prescription continues the kernel in $P$, differentiates
it by $(4/\pi^2)\partial_P$, and then evaluates it at $P=1$.
This derivative converts the continued momentum factor
$\cosh(\pi k\sqrt P)\,dk/\pi$ into the disk measure
$(2/\pi^2)k\sinh(\pi k)\,dk$~\cite{StanfordWitten2017,MertensTuriaciVerlinde2017}.

To obtain the action on the circle, we evaluate $\mathcal S$ on $x_f$.
Because $f'$ and $f''$ are periodic,
$x_f'=\tfrac12 f''/f'+i\pi f'$ has equal values at $u=0$ and $u=1$,
so the endpoint term vanishes. Moreover, $\zeta_f$ is related to
$\tan(\pi f)$ by a M\"obius transformation, which leaves the
Schwarzian derivative invariant. Thus
\begin{equation}
 S_{\rm Sch}[f]=\mathcal S[x_f]
 =-\frac{1}{8I}\int_0^1du\,\{\tan(\pi f(u)),u\},
\label{eq:app-Sch-circle}
\end{equation}
which is Eq.~\eqref{eq:Sch-circle}.

\subsection{Future boundary value}
The future orientation $L=-2iT$ gives a clock length
$I=-2iY$ with $Y>0$.
For $P_{\rm e}=-\ell^2/\pi^2$, the ratio of $A_0$ to the zero-displacement
kernel is $e^{iP_{\rm e}\pi^2/(8Y)}$. Fourier superposition over real
$P_{\rm e}$ with $e^{-i\Phi_FP_{\rm e}}$ selects $Y=\pi^2/(8\Phi_F)$. The
positive $P_{\rm e}$ half uses regulated analytic endpoint displacements.
The resulting future boundary value satisfies the CJ relation
$4I_F=-\pi^2\tau_F$.

\subsection{MPS representation}
We represent the MPS height state $|r-1\rangle$ by
$e_r(\theta)=\sqrt r\,e^{ir\theta}$, with $r\geq1$. The Dirichlet basis on
the disk is $z^r/\sqrt r$, with $r=n+1$, and $e_r$ is its boundary
current. Under the circle
reparametrization $F(\theta)=2\pi f(\theta/2\pi)$, the current pulls back
to $F'(\theta)e_r(F(\theta))$, which we expand in the Fourier
basis~\cite{NagSullivan1995}
\begin{equation}
 F'(\theta)e_r(F(\theta))
 =\sum_{m\geq1}(\alpha_{mr}e_m+\beta_{mr}\bar e_m).
\end{equation}
The coefficient $\beta_{mr}$ represents the negative-frequency
mode $-m$. M\"obius maps preserve the positive modes.
If the mixing vanishes for $r=1$, then $e^{iF}$ is a degree-one
Blaschke map.

On the height basis, the generators of $PSL(2,\mathbb{R})$ act as~\cite{Bargmann1947}
 \begin{align}
 J_0|n\rangle &= (n+1)|n\rangle,\quad
 J_+|n\rangle = \sqrt{(n+1)(n+2)}\,|n+1\rangle,\nonumber\\
 J_-|n\rangle &= \sqrt{n(n+1)}\,|n-1\rangle \quad (n\geq1),
 \qquad J_-|0\rangle=0,
 \end{align}
where $J_0=R$, $J_+=\sqrt{R}S\sqrt{R}$ and $J_+^\dagger=J_-$.
Since $n=0$ corresponds to $r=1$, the lowering action stops at the
lowest positive frequency.
Summing the positive-frequency modes gives the two-point kernel
\begin{equation}
 C(\theta,\vartheta)
 =\sum_{r\geq1}e_r(\theta)\bar e_r(\vartheta)
 =-\frac{1}{4\sin^2[(\theta-\vartheta+i0)/2]} .
\end{equation}

To see how a circle reparametrization changes the two-point kernel,
we replace each mode $e_r(\theta)$ by $F'(\theta)e_r(F(\theta))$.
Summing over the same modes gives
\begin{align}
 C_F(\theta,\vartheta)
 &=F'(\theta)F'(\vartheta)C(F(\theta),F(\vartheta)) \nonumber\\
 &=-\frac{z'(\theta)z'(\vartheta)}
          {[z(\theta)-z(\vartheta)]^2},
 \quad z(\theta)=\tan\!\frac{F(\theta)}{2},
\end{align}
at separated points. The last expression is invariant under
M\"obius transformations of $z$.

For a general $F$, we evaluate the change by bringing the two
points together symmetrically, with
$\theta_\pm=\theta\pm\epsilon/2$. Expanding the numerator and
denominator gives
\begin{equation}
 C_F(\theta_+,\theta_-)
 =-\frac{1}{\epsilon^2}
  -\frac{1}{6}\{z,\theta\}+O(\epsilon^2).
\end{equation}
The reference kernel has the same pole and corresponds to
$z=\tan(\theta/2)$, whose Schwarzian is $1/2$. The subtraction
therefore isolates the finite change~\cite{BelavinPolyakovZamolodchikov1984}
\begin{align}
 \Delta C_F(\theta)
 &\equiv \lim_{\epsilon\to0}
 \bigl[C_F(\theta_+,\theta_-)-C(\theta_+,\theta_-)\bigr]
 \nonumber\\
 &=-\frac{1}{6}\left(\{z,\theta\}-\frac12\right).
\end{align}
Using $\theta=2\pi u$ and $F(\theta)=2\pi f(u)$, its integral
reproduces the circle Schwarzian action relative to the identity,
\begin{equation}
 S_{\rm Sch}[f]-S_{\rm Sch}[\mathrm{id}]
 =\frac{3\pi}{2I}\int_0^{2\pi}d\theta\,\Delta C_F(\theta).
\label{eq:app-kernel-schwarzian}
\end{equation}
Preservation of the current symplectic
form gives $2\sum_{m,r}m|\beta_{mr}|^2$ for the same angular average in
the mode basis. Equating the two expressions yields
\begin{equation}
 -\int_0^{2\pi}\!d\theta\,
 \bigl(\{\tan(F/2),\theta\}-\tfrac12\bigr)
 =24\pi\sum_{m,r\ge1}m|\beta_{mr}|^2 .
\end{equation}
The change of variable $k=2s$ puts the disk density in the convention
$s\sinh(2\pi s)\,ds$. The heat factor $e^{-4Is^2}$ fixes the
unit-period Schwarzian coupling $C_{\rm Sch}=1/(8I)$ through
$e^{-s^2/(2C_{\rm Sch})}$.
In angular units, the coefficient is $2\pi C_{\rm Sch}=\pi/(4I)$, which
gives Eq.~\eqref{eq:Sch-circle} at $I=I_F$.
The identity evaluates smooth boundary configurations. The open kernels
give the quantum distance moments. In the original height
representation, the rotation generator $J_0=R$
obeys $[R,W]=S-S^\dagger$, so the action of $PSL(2,\mathbb R)$ does not
commute with a step of the channel.
\subsection{Static time on the height space}
\label{app:static-height-evolution}

The static coordinates obey
$t_{\rm st}=\tfrac12\log(-z_u z_v)$ and
$r=(z_u+z_v)/(z_v-z_u)$ in the patch $z_u<0<z_v$.
Thus $(z_u,z_v)\mapsto(e^s z_u,e^s z_v)$ advances $t_{\rm st}$ by $s$
at fixed $r$. In the circle representation $z=\tan(\theta/2)$,
this dilation lifts to $F_s$ with
$\tan(F_s(\theta)/2)=e^s\tan(\theta/2)$ and
$\partial_sF_s|_{s=0}=\sin\theta$.
Differentiating the current pullback
$(U_{\rm st}(s)e)(\theta)=F_s'(\theta)e(F_s(\theta))$ gives
\begin{equation}
 \left.\partial_sU_{\rm st}(s)e\right|_{s=0}
 =\partial_\theta(\sin\theta\,e)
 =\tfrac12(J_+-J_-)e .
\end{equation}
Here $J_\pm e=-i\partial_\theta(e^{\pm i\theta}e)$, in agreement
with their action on $|n\rangle$ above. Hence
$U_{\rm st}(s)=e^{-isB_{\rm st}}$, with
\begin{equation}
 B_{\rm st}=\tfrac i2(J_+-J_-)
 =\tfrac i2\sqrt R\,[R,W]\,\sqrt R .
 \label{eq:height-static-generator}
\end{equation}

In fusion SYK, two nested Majorana probe pairs give the two height
tags $D(w)=\sum_{n\ge0}e^{-wn}P_n$, where
$P_n=|n\rangle\langle n|$. Resolving the heights on either side of
one factor of $W$ gives
\begin{equation}
 \langle m|W|n\rangle
 =\langle0|W^m P_m W P_n W^n|0\rangle .
\end{equation}
The exterior factors follow from
$\langle n|W^n|0\rangle=1$, since $n$ upward steps are the
only path from $0$ to $n$ in $n$ steps.
Equation~\eqref{eq:height-static-generator} then weights each directed
step by $\tfrac i2(m-n)\sqrt{(m+1)(n+1)}$.
The probe identity is taken at fixed $m,n$ in the microscopic fusion
limit. It determines the matrix elements of the height-space
generator.
\subsection{Bunch--Davies Wightman functions}
\label{app:height-bd-wightman}

For the two chiralities, the height kets at $\theta$ have
components $e_{n+1}(\theta)$ and $\bar e_{n+1}(\theta)$,
respectively. Their ordered overlaps, with Abel regularization, are
\begin{align}
 C_\pm^+(\theta,\vartheta)
 &=\lim_{\epsilon\downarrow0}\sum_{k=1}^\infty
       k e^{\mp ik(\theta-\vartheta)-\epsilon k}\nonumber\\
 &=-\frac{1}{4\sin^2[(\theta-\vartheta\mp i0)/2]} .
\end{align}
In the projective coordinate $y=\tan(\theta/2)$, the weight-one
transformation and the canonical scalar normalization give
$-1/[4\pi(y-y'\mp i0)^2]$. In the static patch, the two kernels
use $y=z_v=e^{\xi_+}$ and $y=z_u=-e^{\xi_-}$, respectively,
where $\xi_\pm=t_{\rm st}\pm r_*$. In the antiholomorphic sector,
the change from $z_u$ to the future-increasing coordinate
$-z_u=e^{\xi_-}$ turns $+i0$ into $-i0$. Including the Jacobian
at each insertion gives the same form in both null coordinates,
\begin{equation}
 G^+_{{\rm MPS},\pm}(\xi_\pm,\xi_\pm')
 =-\frac{1}{16\pi\sinh^2[(\xi_\pm-\xi_\pm'-i0)/2]} .
 \label{eq:height-bd-wightman}
\end{equation}

For the bulk comparison, the Bunch--Davies two-point function of a
massive scalar on $\mathrm{dS}_2$ has a divergent constant in the
massless limit. Its position-dependent finite term is
$-\log(1-Z_{\rm BD})/(4\pi)$~\cite{BertolaCorbettaMoschella2007}.
Here $Z_{\rm BD}$ is the Wightman boundary value of the invariant
embedding product $Z$, normalized to one at coincidence. In the
coordinates $(z,\tilde z)=(z_v,-z_u)$,
\begin{equation}
 1-Z=-\frac{2(z-z')(\tilde z-\tilde z')}
              {(z+\tilde z)(z'+\tilde z')} .
\end{equation}
The prescription $t_{\rm st}\to t_{\rm st}-i0$ gives $-i0$ in
both $z-z'$ and $\tilde z-\tilde z'$. Differentiating at both
insertions in the same null coordinate retains only the logarithm
of the corresponding difference. With the two Jacobians, this
gives Eq.~\eqref{eq:height-bd-wightman} for each current
$j_\pm=\partial_{\xi_\pm}\phi$.

The mixed derivatives vanish because the logarithm separates into
terms involving a single null difference or a single insertion.
Since $\partial_{t_{\rm st}}\phi=j_++j_-$, the bulk Wightman
function at two points $p$ and $p'$ is therefore
\begin{align}
 &\bigl\langle\partial_{t_{\rm st}}\phi(p)
             \,\partial_{t'_{\rm st}}\phi(p')\bigr\rangle_{\rm BD}
 \nonumber\\
 &\qquad=\sum_{\chi=\pm}
 G^+_{{\rm MPS},\chi}\bigl(\xi_\chi(p),\xi_\chi(p')\bigr).
\end{align}
For each chirality separately, the height ket $|n\rangle$ represents
the one-particle excitation $a^\dagger_{n+1,\pm}|\Omega\rangle$
above the corresponding free-current vacuum.

\subsection{Global conformal action on the heights}
\label{app:global-height-action}

For a real one-parameter projective map $M_s$ with circle lift $F_s$,
set $v(\theta)=\partial_sF_s(\theta)|_{s=0}$. In the convention
$U_M(s)=e^{-isB}$ for the current pullback, the generator is
$B e=i\partial_\theta(v e)$. The translations, dilations, and
special conformal transformations have circle vector fields
$1+\cos\theta$, $\sin\theta$, and $1-\cos\theta$, respectively.
Their generators on the height basis are
\begin{align}
 z\mapsto z+s:\quad
 &B=-J_0-\tfrac12(J_++J_-),\nonumber\\
 z\mapsto e^s z:\quad
 &B=\tfrac i2(J_+-J_-),\nonumber\\
 z\mapsto z/(1-sz):\quad
 &B=-J_0+\tfrac12(J_++J_-).
\end{align}
Thus $J_0$ supplies a height-dependent phase and $J_\pm$ shift the
fusion spin $j=n/2$ by $\pm\tfrac12$.

A finite M\"obius transformation $M$ sends the bulk point
$(z_u,z_v)$ to $(M(z_u),M(z_v))$. With the observer held fixed and
$M(z_u)<0<M(z_v)$, its new static coordinates are
\begin{align}
 t'_{\rm st}&=\tfrac12\log[-M(z_u)M(z_v)],\nonumber\\
 r'&=\frac{M(z_u)+M(z_v)}{M(z_v)-M(z_u)}.
\end{align}
A generic transformation changes both coordinates. The dilation
has $t'_{\rm st}=t_{\rm st}+s$ and $r'=r$.
\section{WZW extension and transport of fusion data}
\label{app:wzw-transport}
\label{supp:wzw}

At $Q=1$, let $W_j$ be the spin-$j$ module of the
$\mathrm{SU}(2)_k$ Wess--Zumino--Witten model and let $P_j$ project
onto its primary multiplet $V_j=W_j[0]$. For each allowed transition
$J=j\pm\tfrac12$ with $0\le j,J\le k/2$, normalize the spin-$1/2$
chiral vertex operator $\Phi_\alpha^{Jj}$ so that its matrix elements
between primary states are the ordinary $\mathrm{SU}(2)$
Clebsch--Gordan coefficients~\cite{KnizhnikZamolodchikov1984,McRae2020}.
The corresponding block of the Markov-reference tensor is then
\begin{equation}
 (B_\alpha)_{Jj}
 =\frac{1}{\sqrt{2}}P_J\Phi_\alpha^{Jj}(1)P_j .
\end{equation}

At $Q>1$, the fusion tensors use $U_Q(\mathfrak{su}(2))$
Clebsch--Gordan coefficients. For the role of quantum groups in WZW
conformal blocks, see Refs.~\cite{Kohno1987,Gaberdiel1994}.

\bibliography{11121545}

\begin{thebibliography}{84}%
\makeatletter
\providecommand \@ifxundefined [1]{%
 \@ifx{#1\undefined}
}%
\providecommand \@ifnum [1]{%
 \ifnum #1\expandafter \@firstoftwo
 \else \expandafter \@secondoftwo
 \fi
}%
\providecommand \@ifx [1]{%
 \ifx #1\expandafter \@firstoftwo
 \else \expandafter \@secondoftwo
 \fi
}%
\providecommand \natexlab [1]{#1}%
\providecommand \enquote  [1]{``#1''}%
\providecommand \bibnamefont  [1]{#1}%
\providecommand \bibfnamefont [1]{#1}%
\providecommand \citenamefont [1]{#1}%
\providecommand \href@noop [0]{\@secondoftwo}%
\providecommand \href [0]{\begingroup \@sanitize@url \@href}%
\providecommand \@href[1]{\@@startlink{#1}\@@href}%
\providecommand \@@href[1]{\endgroup#1\@@endlink}%
\providecommand \@sanitize@url [0]{\catcode `\\12\catcode `\$12\catcode
  `\&12\catcode `\#12\catcode `\^12\catcode `\_12\catcode `\%12\relax}%
\providecommand \@@startlink[1]{}%
\providecommand \@@endlink[0]{}%
\providecommand \url  [0]{\begingroup\@sanitize@url \@url }%
\providecommand \@url [1]{\endgroup\@href {#1}{\urlprefix }}%
\providecommand \urlprefix  [0]{URL }%
\providecommand \Eprint [0]{\href }%
\providecommand \doibase [0]{https://doi.org/}%
\providecommand \selectlanguage [0]{\@gobble}%
\providecommand \bibinfo  [0]{\@secondoftwo}%
\providecommand \bibfield  [0]{\@secondoftwo}%
\providecommand \translation [1]{[#1]}%
\providecommand \BibitemOpen [0]{}%
\providecommand \bibitemStop [0]{}%
\providecommand \bibitemNoStop [0]{.\EOS\space}%
\providecommand \EOS [0]{\spacefactor3000\relax}%
\providecommand \BibitemShut  [1]{\csname bibitem#1\endcsname}%
\let\auto@bib@innerbib\@empty
\bibitem [{\citenamefont {'t~Hooft}(1993)}]{tHooft1993}%
  \BibitemOpen
  \bibfield  {author} {\bibinfo {author} {\bibfnamefont {G.}~\bibnamefont
  {'t~Hooft}},\ }\href@noop {} {\bibinfo {title} {Dimensional reduction in
  quantum gravity}} (\bibinfo {year} {1993}),\ \Eprint
  {https://arxiv.org/abs/gr-qc/9310026} {arXiv:gr-qc/9310026} \BibitemShut
  {NoStop}%
\bibitem [{\citenamefont {Susskind}(1995)}]{Susskind1995}%
  \BibitemOpen
  \bibfield  {author} {\bibinfo {author} {\bibfnamefont {L.}~\bibnamefont
  {Susskind}},\ }\href {https://doi.org/10.1063/1.531249} {\bibfield  {journal}
  {\bibinfo  {journal} {J. Math. Phys.}\ }\textbf {\bibinfo {volume} {36}},\
  \bibinfo {pages} {6377} (\bibinfo {year} {1995})}\BibitemShut {NoStop}%
\bibitem [{\citenamefont {Maldacena}(1998)}]{Maldacena1998}%
  \BibitemOpen
  \bibfield  {author} {\bibinfo {author} {\bibfnamefont {J.~M.}\ \bibnamefont
  {Maldacena}},\ }\href@noop {} {\bibfield  {journal} {\bibinfo  {journal}
  {Adv. Theor. Math. Phys.}\ }\textbf {\bibinfo {volume} {2}},\ \bibinfo
  {pages} {231} (\bibinfo {year} {1998})}\BibitemShut {NoStop}%
\bibitem [{\citenamefont {Gubser}\ \emph {et~al.}(1998)\citenamefont {Gubser},
  \citenamefont {Klebanov},\ and\ \citenamefont
  {Polyakov}}]{GubserKlebanovPolyakov1998}%
  \BibitemOpen
  \bibfield  {author} {\bibinfo {author} {\bibfnamefont {S.~S.}\ \bibnamefont
  {Gubser}}, \bibinfo {author} {\bibfnamefont {I.~R.}\ \bibnamefont
  {Klebanov}},\ and\ \bibinfo {author} {\bibfnamefont {A.~M.}\ \bibnamefont
  {Polyakov}},\ }\href {https://doi.org/10.1016/S0370-2693(98)00377-3}
  {\bibfield  {journal} {\bibinfo  {journal} {Phys. Lett. B}\ }\textbf
  {\bibinfo {volume} {428}},\ \bibinfo {pages} {105} (\bibinfo {year}
  {1998})}\BibitemShut {NoStop}%
\bibitem [{\citenamefont {Witten}(1998)}]{Witten1998Holography}%
  \BibitemOpen
  \bibfield  {author} {\bibinfo {author} {\bibfnamefont {E.}~\bibnamefont
  {Witten}},\ }\href@noop {} {\bibfield  {journal} {\bibinfo  {journal} {Adv.
  Theor. Math. Phys.}\ }\textbf {\bibinfo {volume} {2}},\ \bibinfo {pages}
  {253} (\bibinfo {year} {1998})}\BibitemShut {NoStop}%
\bibitem [{\citenamefont {Teitelboim}(1983)}]{Teitelboim1983}%
  \BibitemOpen
  \bibfield  {author} {\bibinfo {author} {\bibfnamefont {C.}~\bibnamefont
  {Teitelboim}},\ }\href {https://doi.org/10.1016/0370-2693(83)90012-6}
  {\bibfield  {journal} {\bibinfo  {journal} {Phys. Lett. B}\ }\textbf
  {\bibinfo {volume} {126}},\ \bibinfo {pages} {41} (\bibinfo {year}
  {1983})}\BibitemShut {NoStop}%
\bibitem [{\citenamefont {Jackiw}(1985)}]{Jackiw1985}%
  \BibitemOpen
  \bibfield  {author} {\bibinfo {author} {\bibfnamefont {R.}~\bibnamefont
  {Jackiw}},\ }\href {https://doi.org/10.1016/0550-3213(85)90448-1} {\bibfield
  {journal} {\bibinfo  {journal} {Nucl. Phys. B}\ }\textbf {\bibinfo {volume}
  {252}},\ \bibinfo {pages} {343} (\bibinfo {year} {1985})}\BibitemShut
  {NoStop}%
\bibitem [{\citenamefont {Saad}\ \emph {et~al.}(2019)\citenamefont {Saad},
  \citenamefont {Shenker},\ and\ \citenamefont
  {Stanford}}]{SaadShenkerStanford2019}%
  \BibitemOpen
  \bibfield  {author} {\bibinfo {author} {\bibfnamefont {P.}~\bibnamefont
  {Saad}}, \bibinfo {author} {\bibfnamefont {S.~H.}\ \bibnamefont {Shenker}},\
  and\ \bibinfo {author} {\bibfnamefont {D.}~\bibnamefont {Stanford}},\
  }\href@noop {} {\bibinfo {title} {{JT} gravity as a matrix integral}}
  (\bibinfo {year} {2019}),\ \Eprint {https://arxiv.org/abs/1903.11115}
  {arXiv:1903.11115 [hep-th]} \BibitemShut {NoStop}%
\bibitem [{\citenamefont {Sachdev}\ and\ \citenamefont
  {Ye}(1993)}]{SachdevYe1993}%
  \BibitemOpen
  \bibfield  {author} {\bibinfo {author} {\bibfnamefont {S.}~\bibnamefont
  {Sachdev}}\ and\ \bibinfo {author} {\bibfnamefont {J.}~\bibnamefont {Ye}},\
  }\href {https://doi.org/10.1103/PhysRevLett.70.3339} {\bibfield  {journal}
  {\bibinfo  {journal} {Phys. Rev. Lett.}\ }\textbf {\bibinfo {volume} {70}},\
  \bibinfo {pages} {3339} (\bibinfo {year} {1993})}\BibitemShut {NoStop}%
\bibitem [{\citenamefont {Kitaev}(2015)}]{Kitaev2015}%
  \BibitemOpen
  \bibfield  {author} {\bibinfo {author} {\bibfnamefont {A.}~\bibnamefont
  {Kitaev}},\ }\href {https://online.kitp.ucsb.edu/online/entangled15/kitaev/}
  {\bibinfo {title} {A simple model of quantum holography}},\ \bibinfo
  {howpublished} {Talks at the Kavli Institute for Theoretical Physics}
  (\bibinfo {year} {2015})\BibitemShut {NoStop}%
\bibitem [{\citenamefont {Maldacena}\ and\ \citenamefont
  {Stanford}(2016)}]{MaldacenaStanford2016}%
  \BibitemOpen
  \bibfield  {author} {\bibinfo {author} {\bibfnamefont {J.}~\bibnamefont
  {Maldacena}}\ and\ \bibinfo {author} {\bibfnamefont {D.}~\bibnamefont
  {Stanford}},\ }\href {https://doi.org/10.1103/PhysRevD.94.106002} {\bibfield
  {journal} {\bibinfo  {journal} {Phys. Rev. D}\ }\textbf {\bibinfo {volume}
  {94}},\ \bibinfo {pages} {106002} (\bibinfo {year} {2016})}\BibitemShut
  {NoStop}%
\bibitem [{\citenamefont {Witten}(2001)}]{Witten2001dS}%
  \BibitemOpen
  \bibfield  {author} {\bibinfo {author} {\bibfnamefont {E.}~\bibnamefont
  {Witten}},\ }\href@noop {} {\bibinfo {title} {Quantum gravity in de {Sitter}
  space}} (\bibinfo {year} {2001}),\ \Eprint
  {https://arxiv.org/abs/hep-th/0106109} {arXiv:hep-th/0106109} \BibitemShut
  {NoStop}%
\bibitem [{\citenamefont {Strominger}(2001)}]{Strominger2001}%
  \BibitemOpen
  \bibfield  {author} {\bibinfo {author} {\bibfnamefont {A.}~\bibnamefont
  {Strominger}},\ }\href {https://doi.org/10.1088/1126-6708/2001/10/034}
  {\bibfield  {journal} {\bibinfo  {journal} {JHEP}\ }\textbf {\bibinfo
  {volume} {2001}}\bibinfo  {number} { (10)},\ \bibinfo {pages}
  {034}}\BibitemShut {NoStop}%
\bibitem [{\citenamefont {Anninos}\ \emph {et~al.}(2017)\citenamefont
  {Anninos}, \citenamefont {Hartman},\ and\ \citenamefont
  {Strominger}}]{AnninosHartmanStrominger2011}%
  \BibitemOpen
\bibfield  {number} {  }\bibfield  {author} {\bibinfo {author} {\bibfnamefont
  {D.}~\bibnamefont {Anninos}}, \bibinfo {author} {\bibfnamefont
  {T.}~\bibnamefont {Hartman}},\ and\ \bibinfo {author} {\bibfnamefont
  {A.}~\bibnamefont {Strominger}},\ }\href
  {https://doi.org/10.1088/1361-6382/34/1/015009} {\bibfield  {journal}
  {\bibinfo  {journal} {Class. Quantum Grav.}\ }\textbf {\bibinfo {volume}
  {34}},\ \bibinfo {pages} {015009} (\bibinfo {year} {2017})}\BibitemShut
  {NoStop}%
\bibitem [{\citenamefont {Hartle}\ and\ \citenamefont
  {Hawking}(1983)}]{HartleHawking1983}%
  \BibitemOpen
  \bibfield  {author} {\bibinfo {author} {\bibfnamefont {J.~B.}\ \bibnamefont
  {Hartle}}\ and\ \bibinfo {author} {\bibfnamefont {S.~W.}\ \bibnamefont
  {Hawking}},\ }\href {https://doi.org/10.1103/PhysRevD.28.2960} {\bibfield
  {journal} {\bibinfo  {journal} {Phys. Rev. D}\ }\textbf {\bibinfo {volume}
  {28}},\ \bibinfo {pages} {2960} (\bibinfo {year} {1983})}\BibitemShut
  {NoStop}%
\bibitem [{\citenamefont {Maldacena}(2003)}]{Maldacena2003}%
  \BibitemOpen
  \bibfield  {author} {\bibinfo {author} {\bibfnamefont {J.}~\bibnamefont
  {Maldacena}},\ }\href {https://doi.org/10.1088/1126-6708/2003/05/013}
  {\bibfield  {journal} {\bibinfo  {journal} {JHEP}\ }\textbf {\bibinfo
  {volume} {2003}}\bibinfo  {number} { (05)},\ \bibinfo {pages}
  {013}}\BibitemShut {NoStop}%
\bibitem [{\citenamefont {Maldacena}\ \emph {et~al.}(2021)\citenamefont
  {Maldacena}, \citenamefont {Turiaci},\ and\ \citenamefont
  {Yang}}]{MaldacenaTuriaciYang2021}%
  \BibitemOpen
\bibfield  {number} {  }\bibfield  {author} {\bibinfo {author} {\bibfnamefont
  {J.}~\bibnamefont {Maldacena}}, \bibinfo {author} {\bibfnamefont {G.~J.}\
  \bibnamefont {Turiaci}},\ and\ \bibinfo {author} {\bibfnamefont
  {Z.}~\bibnamefont {Yang}},\ }\href {https://doi.org/10.1007/JHEP01(2021)139}
  {\bibfield  {journal} {\bibinfo  {journal} {JHEP}\ }\textbf {\bibinfo
  {volume} {2021}}\bibinfo  {number} { (01)},\ \bibinfo {pages}
  {139}}\BibitemShut {NoStop}%
\bibitem [{\citenamefont {Cotler}\ \emph {et~al.}(2020)\citenamefont {Cotler},
  \citenamefont {Jensen},\ and\ \citenamefont
  {Maloney}}]{CotlerJensenMaloney2020}%
  \BibitemOpen
\bibfield  {number} {  }\bibfield  {author} {\bibinfo {author} {\bibfnamefont
  {J.}~\bibnamefont {Cotler}}, \bibinfo {author} {\bibfnamefont
  {K.}~\bibnamefont {Jensen}},\ and\ \bibinfo {author} {\bibfnamefont
  {A.}~\bibnamefont {Maloney}},\ }\href
  {https://doi.org/10.1007/JHEP06(2020)048} {\bibfield  {journal} {\bibinfo
  {journal} {JHEP}\ }\textbf {\bibinfo {volume} {2020}}\bibinfo  {number} {
  (06)},\ \bibinfo {pages} {048}}\BibitemShut {NoStop}%
\bibitem [{\citenamefont {Cotler}\ and\ \citenamefont
  {Jensen}(2024)}]{CotlerJensen2024}%
  \BibitemOpen
\bibfield  {number} {  }\bibfield  {author} {\bibinfo {author} {\bibfnamefont
  {J.}~\bibnamefont {Cotler}}\ and\ \bibinfo {author} {\bibfnamefont
  {K.}~\bibnamefont {Jensen}},\ }\href
  {https://doi.org/10.1007/JHEP12(2024)016} {\bibfield  {journal} {\bibinfo
  {journal} {JHEP}\ }\textbf {\bibinfo {volume} {2024}}\bibinfo  {number} {
  (12)},\ \bibinfo {pages} {016}}\BibitemShut {NoStop}%
\bibitem [{\citenamefont {Foss-Feig}\ \emph {et~al.}(2021)\citenamefont
  {Foss-Feig}, \citenamefont {Hayes}, \citenamefont {Dreiling}, \citenamefont
  {Figgatt}, \citenamefont {Gaebler}, \citenamefont {Moses}, \citenamefont
  {Pino},\ and\ \citenamefont {Potter}}]{FossFeigEtAl2021}%
  \BibitemOpen
\bibfield  {number} {  }\bibfield  {author} {\bibinfo {author} {\bibfnamefont
  {M.}~\bibnamefont {Foss-Feig}}, \bibinfo {author} {\bibfnamefont
  {D.}~\bibnamefont {Hayes}}, \bibinfo {author} {\bibfnamefont {J.~M.}\
  \bibnamefont {Dreiling}}, \bibinfo {author} {\bibfnamefont {C.}~\bibnamefont
  {Figgatt}}, \bibinfo {author} {\bibfnamefont {J.~P.}\ \bibnamefont
  {Gaebler}}, \bibinfo {author} {\bibfnamefont {S.~A.}\ \bibnamefont {Moses}},
  \bibinfo {author} {\bibfnamefont {J.~M.}\ \bibnamefont {Pino}},\ and\
  \bibinfo {author} {\bibfnamefont {A.~C.}\ \bibnamefont {Potter}},\ }\href
  {https://doi.org/10.1103/PhysRevResearch.3.033002} {\bibfield  {journal}
  {\bibinfo  {journal} {Phys. Rev. Research}\ }\textbf {\bibinfo {volume}
  {3}},\ \bibinfo {pages} {033002} (\bibinfo {year} {2021})}\BibitemShut
  {NoStop}%
\bibitem [{\citenamefont {Okuyama}(2025{\natexlab{a}})}]{Okuyama2025TN}%
  \BibitemOpen
  \bibfield  {author} {\bibinfo {author} {\bibfnamefont {K.}~\bibnamefont
  {Okuyama}},\ }\href {https://doi.org/10.21468/SciPostPhys.19.4.083}
  {\bibfield  {journal} {\bibinfo  {journal} {SciPost Phys.}\ }\textbf
  {\bibinfo {volume} {19}},\ \bibinfo {pages} {083} (\bibinfo {year}
  {2025}{\natexlab{a}})}\BibitemShut {NoStop}%
\bibitem [{\citenamefont {Narovlansky}\ and\ \citenamefont
  {Verlinde}(2023)}]{NarovlanskyVerlinde2023}%
  \BibitemOpen
  \bibfield  {author} {\bibinfo {author} {\bibfnamefont {V.}~\bibnamefont
  {Narovlansky}}\ and\ \bibinfo {author} {\bibfnamefont {H.}~\bibnamefont
  {Verlinde}},\ }\href@noop {} {\bibinfo {title} {Double-scaled {SYK} and de
  {Sitter} holography}} (\bibinfo {year} {2023}),\ \Eprint
  {https://arxiv.org/abs/2310.16994} {arXiv:2310.16994 [hep-th]} \BibitemShut
  {NoStop}%
\bibitem [{\citenamefont {Okuyama}(2025{\natexlab{b}})}]{Okuyama2025dS}%
  \BibitemOpen
  \bibfield  {author} {\bibinfo {author} {\bibfnamefont {K.}~\bibnamefont
  {Okuyama}},\ }\href {https://doi.org/10.1007/JHEP08(2025)181} {\bibfield
  {journal} {\bibinfo  {journal} {JHEP}\ }\textbf {\bibinfo {volume}
  {2025}}\bibinfo  {number} { (08)},\ \bibinfo {pages} {181}}\BibitemShut
  {NoStop}%
\bibitem [{\citenamefont {Feng}\ \emph {et~al.}(2019)\citenamefont {Feng},
  \citenamefont {Tian},\ and\ \citenamefont {Wei}}]{FengTianWei2019}%
  \BibitemOpen
\bibfield  {number} {  }\bibfield  {author} {\bibinfo {author} {\bibfnamefont
  {R.}~\bibnamefont {Feng}}, \bibinfo {author} {\bibfnamefont {G.}~\bibnamefont
  {Tian}},\ and\ \bibinfo {author} {\bibfnamefont {D.}~\bibnamefont {Wei}},\
  }\href {https://doi.org/10.1007/s42543-018-0007-1} {\bibfield  {journal}
  {\bibinfo  {journal} {Peking Math. J.}\ }\textbf {\bibinfo {volume} {2}},\
  \bibinfo {pages} {41} (\bibinfo {year} {2019})}\BibitemShut {NoStop}%
\bibitem [{\citenamefont {Chen}\ \emph {et~al.}(2017)\citenamefont {Chen},
  \citenamefont {Fradkin},\ and\ \citenamefont
  {Witczak-Krempa}}]{ChenFradkinWitczakKrempa2017}%
  \BibitemOpen
  \bibfield  {author} {\bibinfo {author} {\bibfnamefont {X.}~\bibnamefont
  {Chen}}, \bibinfo {author} {\bibfnamefont {E.}~\bibnamefont {Fradkin}},\ and\
  \bibinfo {author} {\bibfnamefont {W.}~\bibnamefont {Witczak-Krempa}},\ }\href
  {https://doi.org/10.1088/1751-8121/aa8dbc} {\bibfield  {journal} {\bibinfo
  {journal} {J. Phys. A: Math. Theor.}\ }\textbf {\bibinfo {volume} {50}},\
  \bibinfo {pages} {464002} (\bibinfo {year} {2017})}\BibitemShut {NoStop}%
\bibitem [{\citenamefont {Lee}\ and\ \citenamefont {Youn}(2022)}]{LeeYoun2022}%
  \BibitemOpen
  \bibfield  {author} {\bibinfo {author} {\bibfnamefont {H.~H.}\ \bibnamefont
  {Lee}}\ and\ \bibinfo {author} {\bibfnamefont {S.-G.}\ \bibnamefont {Youn}},\
  }\href {https://doi.org/10.1007/s00220-021-04283-9} {\bibfield  {journal}
  {\bibinfo  {journal} {Commun. Math. Phys.}\ }\textbf {\bibinfo {volume}
  {389}},\ \bibinfo {pages} {1303} (\bibinfo {year} {2022})}\BibitemShut
  {NoStop}%
\bibitem [{\citenamefont {Sch{\"o}n}\ \emph {et~al.}(2005)\citenamefont
  {Sch{\"o}n}, \citenamefont {Solano}, \citenamefont {Verstraete},
  \citenamefont {Cirac},\ and\ \citenamefont {Wolf}}]{SchonEtAl2005}%
  \BibitemOpen
  \bibfield  {author} {\bibinfo {author} {\bibfnamefont {C.}~\bibnamefont
  {Sch{\"o}n}}, \bibinfo {author} {\bibfnamefont {E.}~\bibnamefont {Solano}},
  \bibinfo {author} {\bibfnamefont {F.}~\bibnamefont {Verstraete}}, \bibinfo
  {author} {\bibfnamefont {J.~I.}\ \bibnamefont {Cirac}},\ and\ \bibinfo
  {author} {\bibfnamefont {M.~M.}\ \bibnamefont {Wolf}},\ }\href
  {https://doi.org/10.1103/PhysRevLett.95.110503} {\bibfield  {journal}
  {\bibinfo  {journal} {Phys. Rev. Lett.}\ }\textbf {\bibinfo {volume} {95}},\
  \bibinfo {pages} {110503} (\bibinfo {year} {2005})}\BibitemShut {NoStop}%
\bibitem [{\citenamefont {Fannes}\ \emph {et~al.}(1992)\citenamefont {Fannes},
  \citenamefont {Nachtergaele},\ and\ \citenamefont
  {Werner}}]{FannesNachtergaeleWerner1992}%
  \BibitemOpen
  \bibfield  {author} {\bibinfo {author} {\bibfnamefont {M.}~\bibnamefont
  {Fannes}}, \bibinfo {author} {\bibfnamefont {B.}~\bibnamefont
  {Nachtergaele}},\ and\ \bibinfo {author} {\bibfnamefont {R.~F.}\ \bibnamefont
  {Werner}},\ }\href {https://doi.org/10.1007/BF02099178} {\bibfield  {journal}
  {\bibinfo  {journal} {Commun. Math. Phys.}\ }\textbf {\bibinfo {volume}
  {144}},\ \bibinfo {pages} {443} (\bibinfo {year} {1992})}\BibitemShut
  {NoStop}%
\bibitem [{\citenamefont {Cirac}\ \emph {et~al.}(2021)\citenamefont {Cirac},
  \citenamefont {P{\'e}rez-Garc{\'i}a}, \citenamefont {Schuch},\ and\
  \citenamefont {Verstraete}}]{CiracEtAl2021}%
  \BibitemOpen
  \bibfield  {author} {\bibinfo {author} {\bibfnamefont {J.~I.}\ \bibnamefont
  {Cirac}}, \bibinfo {author} {\bibfnamefont {D.}~\bibnamefont
  {P{\'e}rez-Garc{\'i}a}}, \bibinfo {author} {\bibfnamefont {N.}~\bibnamefont
  {Schuch}},\ and\ \bibinfo {author} {\bibfnamefont {F.}~\bibnamefont
  {Verstraete}},\ }\href {https://doi.org/10.1103/RevModPhys.93.045003}
  {\bibfield  {journal} {\bibinfo  {journal} {Rev. Mod. Phys.}\ }\textbf
  {\bibinfo {volume} {93}},\ \bibinfo {pages} {045003} (\bibinfo {year}
  {2021})}\BibitemShut {NoStop}%
\bibitem [{\citenamefont {Karlin}\ and\ \citenamefont
  {McGregor}(1959)}]{KarlinMcGregor1959}%
  \BibitemOpen
  \bibfield  {author} {\bibinfo {author} {\bibfnamefont {S.}~\bibnamefont
  {Karlin}}\ and\ \bibinfo {author} {\bibfnamefont {J.}~\bibnamefont
  {McGregor}},\ }\href {https://doi.org/10.1215/ijm/1255454999} {\bibfield
  {journal} {\bibinfo  {journal} {Illinois J. Math.}\ }\textbf {\bibinfo
  {volume} {3}},\ \bibinfo {pages} {66} (\bibinfo {year} {1959})}\BibitemShut
  {NoStop}%
\bibitem [{\citenamefont {Neshveyev}\ and\ \citenamefont
  {Tuset}(2004)}]{NeshveyevTuset2004}%
  \BibitemOpen
  \bibfield  {author} {\bibinfo {author} {\bibfnamefont {S.}~\bibnamefont
  {Neshveyev}}\ and\ \bibinfo {author} {\bibfnamefont {L.}~\bibnamefont
  {Tuset}},\ }\href {https://doi.org/10.1515/crll.2004.018} {\bibfield
  {journal} {\bibinfo  {journal} {J. Reine Angew. Math.}\ }\textbf {\bibinfo
  {volume} {568}},\ \bibinfo {pages} {23} (\bibinfo {year} {2004})}\BibitemShut
  {NoStop}%
\bibitem [{\citenamefont {Garc{\'i}a-Garc{\'i}a}\ \emph
  {et~al.}(2018)\citenamefont {Garc{\'i}a-Garc{\'i}a}, \citenamefont {Jia},\
  and\ \citenamefont {Verbaarschot}}]{GarciaGarciaJiaVerbaarschot2018}%
  \BibitemOpen
  \bibfield  {author} {\bibinfo {author} {\bibfnamefont {A.~M.}\ \bibnamefont
  {Garc{\'i}a-Garc{\'i}a}}, \bibinfo {author} {\bibfnamefont {Y.}~\bibnamefont
  {Jia}},\ and\ \bibinfo {author} {\bibfnamefont {J.~J.~M.}\ \bibnamefont
  {Verbaarschot}},\ }\href {https://doi.org/10.1007/JHEP04(2018)146} {\bibfield
   {journal} {\bibinfo  {journal} {JHEP}\ }\textbf {\bibinfo {volume}
  {2018}}\bibinfo  {number} { (04)},\ \bibinfo {pages} {146}}\BibitemShut
  {NoStop}%
\bibitem [{\citenamefont {Berkooz}\ \emph {et~al.}(2019)\citenamefont
  {Berkooz}, \citenamefont {Isachenkov}, \citenamefont {Narovlansky},\ and\
  \citenamefont {Torrents}}]{BINT2019}%
  \BibitemOpen
\bibfield  {number} {  }\bibfield  {author} {\bibinfo {author} {\bibfnamefont
  {M.}~\bibnamefont {Berkooz}}, \bibinfo {author} {\bibfnamefont
  {M.}~\bibnamefont {Isachenkov}}, \bibinfo {author} {\bibfnamefont
  {V.}~\bibnamefont {Narovlansky}},\ and\ \bibinfo {author} {\bibfnamefont
  {G.}~\bibnamefont {Torrents}},\ }\href
  {https://doi.org/10.1007/JHEP03(2019)079} {\bibfield  {journal} {\bibinfo
  {journal} {JHEP}\ }\textbf {\bibinfo {volume} {2019}}\bibinfo  {number} {
  (03)},\ \bibinfo {pages} {079}}\BibitemShut {NoStop}%
\bibitem [{\citenamefont {Lin}(2022)}]{Lin2022}%
  \BibitemOpen
\bibfield  {number} {  }\bibfield  {author} {\bibinfo {author} {\bibfnamefont
  {H.~W.}\ \bibnamefont {Lin}},\ }\href
  {https://doi.org/10.1007/JHEP11(2022)060} {\bibfield  {journal} {\bibinfo
  {journal} {JHEP}\ }\textbf {\bibinfo {volume} {2022}}\bibinfo  {number} {
  (11)},\ \bibinfo {pages} {060}}\BibitemShut {NoStop}%
\bibitem [{\citenamefont {Cotler}\ \emph {et~al.}(2017)\citenamefont {Cotler},
  \citenamefont {Gur-Ari}, \citenamefont {Hanada}, \citenamefont {Polchinski},
  \citenamefont {Saad}, \citenamefont {Shenker}, \citenamefont {Stanford},
  \citenamefont {Streicher},\ and\ \citenamefont {Tezuka}}]{CotlerEtAl2017}%
  \BibitemOpen
\bibfield  {number} {  }\bibfield  {author} {\bibinfo {author} {\bibfnamefont
  {J.~S.}\ \bibnamefont {Cotler}}, \bibinfo {author} {\bibfnamefont
  {G.}~\bibnamefont {Gur-Ari}}, \bibinfo {author} {\bibfnamefont
  {M.}~\bibnamefont {Hanada}}, \bibinfo {author} {\bibfnamefont
  {J.}~\bibnamefont {Polchinski}}, \bibinfo {author} {\bibfnamefont
  {P.}~\bibnamefont {Saad}}, \bibinfo {author} {\bibfnamefont {S.~H.}\
  \bibnamefont {Shenker}}, \bibinfo {author} {\bibfnamefont {D.}~\bibnamefont
  {Stanford}}, \bibinfo {author} {\bibfnamefont {A.}~\bibnamefont
  {Streicher}},\ and\ \bibinfo {author} {\bibfnamefont {M.}~\bibnamefont
  {Tezuka}},\ }\href {https://doi.org/10.1007/JHEP05(2017)118} {\bibfield
  {journal} {\bibinfo  {journal} {JHEP}\ }\textbf {\bibinfo {volume}
  {2017}}\bibinfo  {number} { (05)},\ \bibinfo {pages} {118}}\BibitemShut
  {NoStop}%
\bibitem [{\citenamefont {Ismail}\ \emph {et~al.}(1987)\citenamefont {Ismail},
  \citenamefont {Stanton},\ and\ \citenamefont
  {Viennot}}]{IsmailStantonViennot1987}%
  \BibitemOpen
\bibfield  {number} {  }\bibfield  {author} {\bibinfo {author} {\bibfnamefont
  {M.~E.~H.}\ \bibnamefont {Ismail}}, \bibinfo {author} {\bibfnamefont
  {D.}~\bibnamefont {Stanton}},\ and\ \bibinfo {author} {\bibfnamefont
  {G.}~\bibnamefont {Viennot}},\ }\href
  {https://doi.org/10.1016/S0195-6698(87)80046-X} {\bibfield  {journal}
  {\bibinfo  {journal} {Eur. J. Comb.}\ }\textbf {\bibinfo {volume} {8}},\
  \bibinfo {pages} {379} (\bibinfo {year} {1987})}\BibitemShut {NoStop}%
\bibitem [{\citenamefont {Rabinovici}\ \emph {et~al.}(2023)\citenamefont
  {Rabinovici}, \citenamefont {S{\'a}nchez-Garrido}, \citenamefont {Shir},\
  and\ \citenamefont {Sonner}}]{RabinoviciEtAl2023}%
  \BibitemOpen
  \bibfield  {author} {\bibinfo {author} {\bibfnamefont {E.}~\bibnamefont
  {Rabinovici}}, \bibinfo {author} {\bibfnamefont {A.}~\bibnamefont
  {S{\'a}nchez-Garrido}}, \bibinfo {author} {\bibfnamefont {R.}~\bibnamefont
  {Shir}},\ and\ \bibinfo {author} {\bibfnamefont {J.}~\bibnamefont {Sonner}},\
  }\href {https://doi.org/10.1007/JHEP08(2023)213} {\bibfield  {journal}
  {\bibinfo  {journal} {JHEP}\ }\textbf {\bibinfo {volume} {2023}}\bibinfo
  {number} { (08)},\ \bibinfo {pages} {213}}\BibitemShut {NoStop}%
\bibitem [{\citenamefont {Berkooz}\ \emph {et~al.}(2023)\citenamefont
  {Berkooz}, \citenamefont {Isachenkov}, \citenamefont {Narayan},\ and\
  \citenamefont {Narovlansky}}]{BerkoozIsachenkovNarayanNarovlansky2023}%
  \BibitemOpen
\bibfield  {number} {  }\bibfield  {author} {\bibinfo {author} {\bibfnamefont
  {M.}~\bibnamefont {Berkooz}}, \bibinfo {author} {\bibfnamefont
  {M.}~\bibnamefont {Isachenkov}}, \bibinfo {author} {\bibfnamefont
  {P.}~\bibnamefont {Narayan}},\ and\ \bibinfo {author} {\bibfnamefont
  {V.}~\bibnamefont {Narovlansky}},\ }\href
  {https://doi.org/10.1007/JHEP08(2023)076} {\bibfield  {journal} {\bibinfo
  {journal} {JHEP}\ }\textbf {\bibinfo {volume} {2023}}\bibinfo  {number} {
  (08)},\ \bibinfo {pages} {076}}\BibitemShut {NoStop}%
\bibitem [{\citenamefont {Blommaert}\ \emph
  {et~al.}(2024{\natexlab{a}})\citenamefont {Blommaert}, \citenamefont
  {Mertens},\ and\ \citenamefont {Yao}}]{BlommaertMertensYao2024Actions}%
  \BibitemOpen
\bibfield  {number} {  }\bibfield  {author} {\bibinfo {author} {\bibfnamefont
  {A.}~\bibnamefont {Blommaert}}, \bibinfo {author} {\bibfnamefont {T.~G.}\
  \bibnamefont {Mertens}},\ and\ \bibinfo {author} {\bibfnamefont
  {S.}~\bibnamefont {Yao}},\ }\href {https://doi.org/10.1007/JHEP02(2024)067}
  {\bibfield  {journal} {\bibinfo  {journal} {JHEP}\ }\textbf {\bibinfo
  {volume} {2024}}\bibinfo  {number} { (02)},\ \bibinfo {pages}
  {067}}\BibitemShut {NoStop}%
\bibitem [{\citenamefont {Liu}\ and\ \citenamefont {Shen}(2026)}]{LiuShen2026}%
  \BibitemOpen
\bibfield  {number} {  }\bibfield  {author} {\bibinfo {author} {\bibfnamefont
  {W.}~\bibnamefont {Liu}}\ and\ \bibinfo {author} {\bibfnamefont
  {H.}~\bibnamefont {Shen}},\ }\href@noop {} {\bibinfo {title} {Limit joint
  distributions of {SYK} models with partial interactions, mixed {$q$}-gaussian
  models and asymptotic {$\varepsilon$}-freeness}} (\bibinfo {year} {2026}),\
  \Eprint {https://arxiv.org/abs/2602.00789} {arXiv:2602.00789 [math.OA]}
  \BibitemShut {NoStop}%
\bibitem [{\citenamefont {Schaller}\ and\ \citenamefont
  {Strobl}(1994)}]{SchallerStrobl1994}%
  \BibitemOpen
  \bibfield  {author} {\bibinfo {author} {\bibfnamefont {P.}~\bibnamefont
  {Schaller}}\ and\ \bibinfo {author} {\bibfnamefont {T.}~\bibnamefont
  {Strobl}},\ }\href {https://doi.org/10.1142/S0217732394002951} {\bibfield
  {journal} {\bibinfo  {journal} {Mod. Phys. Lett. A}\ }\textbf {\bibinfo
  {volume} {9}},\ \bibinfo {pages} {3129} (\bibinfo {year} {1994})}\BibitemShut
  {NoStop}%
\bibitem [{\citenamefont {Ikeda}(1994)}]{Ikeda1994}%
  \BibitemOpen
  \bibfield  {author} {\bibinfo {author} {\bibfnamefont {N.}~\bibnamefont
  {Ikeda}},\ }\href {https://doi.org/10.1006/aphy.1994.1104} {\bibfield
  {journal} {\bibinfo  {journal} {Annals Phys.}\ }\textbf {\bibinfo {volume}
  {235}},\ \bibinfo {pages} {435} (\bibinfo {year} {1994})}\BibitemShut
  {NoStop}%
\bibitem [{\citenamefont {Grumiller}\ \emph {et~al.}(2002)\citenamefont
  {Grumiller}, \citenamefont {Kummer},\ and\ \citenamefont
  {Vassilevich}}]{GrumillerKummerVassilevich2002}%
  \BibitemOpen
  \bibfield  {author} {\bibinfo {author} {\bibfnamefont {D.}~\bibnamefont
  {Grumiller}}, \bibinfo {author} {\bibfnamefont {W.}~\bibnamefont {Kummer}},\
  and\ \bibinfo {author} {\bibfnamefont {D.~V.}\ \bibnamefont {Vassilevich}},\
  }\href {https://doi.org/10.1016/S0370-1573(02)00267-3} {\bibfield  {journal}
  {\bibinfo  {journal} {Phys. Rept.}\ }\textbf {\bibinfo {volume} {369}},\
  \bibinfo {pages} {327} (\bibinfo {year} {2002})}\BibitemShut {NoStop}%
\bibitem [{\citenamefont {Blommaert}\ \emph {et~al.}(2025)\citenamefont
  {Blommaert}, \citenamefont {Mertens},\ and\ \citenamefont
  {Papalini}}]{BlommaertMertensPapalini2025}%
  \BibitemOpen
  \bibfield  {author} {\bibinfo {author} {\bibfnamefont {A.}~\bibnamefont
  {Blommaert}}, \bibinfo {author} {\bibfnamefont {T.~G.}\ \bibnamefont
  {Mertens}},\ and\ \bibinfo {author} {\bibfnamefont {J.}~\bibnamefont
  {Papalini}},\ }\href {https://doi.org/10.1007/JHEP06(2025)050} {\bibfield
  {journal} {\bibinfo  {journal} {JHEP}\ }\textbf {\bibinfo {volume}
  {2025}}\bibinfo  {number} { (06)},\ \bibinfo {pages} {050}}\BibitemShut
  {NoStop}%
\bibitem [{\citenamefont {Blommaert}\ \emph
  {et~al.}(2024{\natexlab{b}})\citenamefont {Blommaert}, \citenamefont
  {Mertens},\ and\ \citenamefont {Yao}}]{BlommaertMertensYao2024Liouville}%
  \BibitemOpen
\bibfield  {number} {  }\bibfield  {author} {\bibinfo {author} {\bibfnamefont
  {A.}~\bibnamefont {Blommaert}}, \bibinfo {author} {\bibfnamefont {T.~G.}\
  \bibnamefont {Mertens}},\ and\ \bibinfo {author} {\bibfnamefont
  {S.}~\bibnamefont {Yao}},\ }\href {https://doi.org/10.1007/JHEP11(2024)054}
  {\bibfield  {journal} {\bibinfo  {journal} {JHEP}\ }\textbf {\bibinfo
  {volume} {2024}}\bibinfo  {number} { (11)},\ \bibinfo {pages}
  {054}}\BibitemShut {NoStop}%
\bibitem [{\citenamefont {Okuyama}\ and\ \citenamefont
  {Sakai}(2020)}]{OkuyamaSakai2020}%
  \BibitemOpen
\bibfield  {number} {  }\bibfield  {author} {\bibinfo {author} {\bibfnamefont
  {K.}~\bibnamefont {Okuyama}}\ and\ \bibinfo {author} {\bibfnamefont
  {K.}~\bibnamefont {Sakai}},\ }\href {https://doi.org/10.1007/JHEP01(2020)156}
  {\bibfield  {journal} {\bibinfo  {journal} {JHEP}\ }\textbf {\bibinfo
  {volume} {2020}}\bibinfo  {number} { (01)},\ \bibinfo {pages}
  {156}}\BibitemShut {NoStop}%
\bibitem [{\citenamefont {Eynard}\ and\ \citenamefont
  {Orantin}(2007{\natexlab{a}})}]{EynardOrantinTopRec2007}%
  \BibitemOpen
\bibfield  {number} {  }\bibfield  {author} {\bibinfo {author} {\bibfnamefont
  {B.}~\bibnamefont {Eynard}}\ and\ \bibinfo {author} {\bibfnamefont
  {N.}~\bibnamefont {Orantin}},\ }\href
  {https://doi.org/10.4310/CNTP.2007.v1.n2.a4} {\bibfield  {journal} {\bibinfo
  {journal} {Commun. Number Theory Phys.}\ }\textbf {\bibinfo {volume} {1}},\
  \bibinfo {pages} {347} (\bibinfo {year} {2007}{\natexlab{a}})}\BibitemShut
  {NoStop}%
\bibitem [{\citenamefont {Eynard}\ and\ \citenamefont
  {Orantin}(2007{\natexlab{b}})}]{EynardOrantin2007}%
  \BibitemOpen
  \bibfield  {author} {\bibinfo {author} {\bibfnamefont {B.}~\bibnamefont
  {Eynard}}\ and\ \bibinfo {author} {\bibfnamefont {N.}~\bibnamefont
  {Orantin}},\ }\href@noop {} {\bibinfo {title} {{Weil--Petersson} volume of
  moduli spaces, {Mirzakhani}'s recursion and matrix models}} (\bibinfo {year}
  {2007}{\natexlab{b}}),\ \Eprint {https://arxiv.org/abs/0705.3600}
  {arXiv:0705.3600 [math-ph]} \BibitemShut {NoStop}%
\bibitem [{\citenamefont {Johnson}(2020)}]{Johnson2020}%
  \BibitemOpen
  \bibfield  {author} {\bibinfo {author} {\bibfnamefont {C.~V.}\ \bibnamefont
  {Johnson}},\ }\href {https://doi.org/10.1103/PhysRevD.101.106023} {\bibfield
  {journal} {\bibinfo  {journal} {Phys. Rev. D}\ }\textbf {\bibinfo {volume}
  {101}},\ \bibinfo {pages} {106023} (\bibinfo {year} {2020})}\BibitemShut
  {NoStop}%
\bibitem [{\citenamefont {Salberger}\ and\ \citenamefont
  {Korepin}(2016)}]{SalbergerKorepin2016}%
  \BibitemOpen
  \bibfield  {author} {\bibinfo {author} {\bibfnamefont {O.}~\bibnamefont
  {Salberger}}\ and\ \bibinfo {author} {\bibfnamefont {V.}~\bibnamefont
  {Korepin}},\ }\href@noop {} {\bibinfo {title} {Fredkin spin chain}} (\bibinfo
  {year} {2016}),\ \Eprint {https://arxiv.org/abs/1605.03842} {arXiv:1605.03842
  [quant-ph]} \BibitemShut {NoStop}%
\bibitem [{\citenamefont {Maldacena}\ \emph {et~al.}(2016)\citenamefont
  {Maldacena}, \citenamefont {Stanford},\ and\ \citenamefont
  {Yang}}]{MaldacenaStanfordYang2016}%
  \BibitemOpen
  \bibfield  {author} {\bibinfo {author} {\bibfnamefont {J.}~\bibnamefont
  {Maldacena}}, \bibinfo {author} {\bibfnamefont {D.}~\bibnamefont
  {Stanford}},\ and\ \bibinfo {author} {\bibfnamefont {Z.}~\bibnamefont
  {Yang}},\ }\href {https://doi.org/10.1093/ptep/ptw124} {\bibfield  {journal}
  {\bibinfo  {journal} {Prog. Theor. Exp. Phys.}\ }\textbf {\bibinfo {volume}
  {2016}},\ \bibinfo {pages} {12C104} (\bibinfo {year} {2016})}\BibitemShut
  {NoStop}%
\bibitem [{\citenamefont {Mertens}\ \emph {et~al.}(2017)\citenamefont
  {Mertens}, \citenamefont {Turiaci},\ and\ \citenamefont
  {Verlinde}}]{MertensTuriaciVerlinde2017}%
  \BibitemOpen
  \bibfield  {author} {\bibinfo {author} {\bibfnamefont {T.~G.}\ \bibnamefont
  {Mertens}}, \bibinfo {author} {\bibfnamefont {G.~J.}\ \bibnamefont
  {Turiaci}},\ and\ \bibinfo {author} {\bibfnamefont {H.~L.}\ \bibnamefont
  {Verlinde}},\ }\href {https://doi.org/10.1007/JHEP08(2017)136} {\bibfield
  {journal} {\bibinfo  {journal} {JHEP}\ }\textbf {\bibinfo {volume}
  {2017}}\bibinfo  {number} { (08)},\ \bibinfo {pages} {136}}\BibitemShut
  {NoStop}%
\bibitem [{\citenamefont {Losev}(2026)}]{Losev2024}%
  \BibitemOpen
\bibfield  {number} {  }\bibfield  {author} {\bibinfo {author} {\bibfnamefont
  {I.}~\bibnamefont {Losev}},\ }\href {https://doi.org/10.1112/plms.70169}
  {\bibfield  {journal} {\bibinfo  {journal} {Proc. Lond. Math. Soc.}\ }\textbf
  {\bibinfo {volume} {132}},\ \bibinfo {pages} {e70169} (\bibinfo {year}
  {2026})}\BibitemShut {NoStop}%
\bibitem [{\citenamefont {Czech}\ \emph {et~al.}(2015)\citenamefont {Czech},
  \citenamefont {Lamprou}, \citenamefont {McCandlish},\ and\ \citenamefont
  {Sully}}]{CzechLamprouMcCandlishSully2015}%
  \BibitemOpen
  \bibfield  {author} {\bibinfo {author} {\bibfnamefont {B.}~\bibnamefont
  {Czech}}, \bibinfo {author} {\bibfnamefont {L.}~\bibnamefont {Lamprou}},
  \bibinfo {author} {\bibfnamefont {S.}~\bibnamefont {McCandlish}},\ and\
  \bibinfo {author} {\bibfnamefont {J.}~\bibnamefont {Sully}},\ }\href
  {https://doi.org/10.1007/JHEP10(2015)175} {\bibfield  {journal} {\bibinfo
  {journal} {JHEP}\ }\textbf {\bibinfo {volume} {2015}}\bibinfo  {number} {
  (10)},\ \bibinfo {pages} {175}}\BibitemShut {NoStop}%
\bibitem [{\citenamefont {Cirac}\ and\ \citenamefont
  {Sierra}(2010)}]{CiracSierra2010}%
  \BibitemOpen
\bibfield  {number} {  }\bibfield  {author} {\bibinfo {author} {\bibfnamefont
  {J.~I.}\ \bibnamefont {Cirac}}\ and\ \bibinfo {author} {\bibfnamefont
  {G.}~\bibnamefont {Sierra}},\ }\href
  {https://doi.org/10.1103/PhysRevB.81.104431} {\bibfield  {journal} {\bibinfo
  {journal} {Phys. Rev. B}\ }\textbf {\bibinfo {volume} {81}},\ \bibinfo
  {pages} {104431} (\bibinfo {year} {2010})}\BibitemShut {NoStop}%
\bibitem [{\citenamefont {Nielsen}\ \emph {et~al.}(2011)\citenamefont
  {Nielsen}, \citenamefont {Cirac},\ and\ \citenamefont
  {Sierra}}]{NielsenCiracSierra2011}%
  \BibitemOpen
  \bibfield  {author} {\bibinfo {author} {\bibfnamefont {A.~E.~B.}\
  \bibnamefont {Nielsen}}, \bibinfo {author} {\bibfnamefont {J.~I.}\
  \bibnamefont {Cirac}},\ and\ \bibinfo {author} {\bibfnamefont
  {G.}~\bibnamefont {Sierra}},\ }\href
  {https://doi.org/10.1088/1742-5468/2011/11/P11014} {\bibfield  {journal}
  {\bibinfo  {journal} {J. Stat. Mech.}\ }\textbf {\bibinfo {volume} {2011}},\
  \bibinfo {pages} {P11014} (\bibinfo {year} {2011})}\BibitemShut {NoStop}%
\bibitem [{\citenamefont {Franco-Rubio}\ \emph {et~al.}(2025)\citenamefont
  {Franco-Rubio}, \citenamefont {Cirac},\ and\ \citenamefont
  {Sierra}}]{FrancoRubioCiracSierra2025}%
  \BibitemOpen
  \bibfield  {author} {\bibinfo {author} {\bibfnamefont {A.}~\bibnamefont
  {Franco-Rubio}}, \bibinfo {author} {\bibfnamefont {J.~I.}\ \bibnamefont
  {Cirac}},\ and\ \bibinfo {author} {\bibfnamefont {G.}~\bibnamefont
  {Sierra}},\ }\href@noop {} {\bibinfo {title} {Matrix product states as thin
  torus limits of conformal correlators}} (\bibinfo {year} {2025}),\ \Eprint
  {https://arxiv.org/abs/2507.22735} {arXiv:2507.22735 [quant-ph]} \BibitemShut
  {NoStop}%
\bibitem [{\citenamefont {de~Alfaro}\ \emph {et~al.}(1976)\citenamefont
  {de~Alfaro}, \citenamefont {Fubini},\ and\ \citenamefont
  {Furlan}}]{deAlfaroFubiniFurlan1976}%
  \BibitemOpen
  \bibfield  {author} {\bibinfo {author} {\bibfnamefont {V.}~\bibnamefont
  {de~Alfaro}}, \bibinfo {author} {\bibfnamefont {S.}~\bibnamefont {Fubini}},\
  and\ \bibinfo {author} {\bibfnamefont {G.}~\bibnamefont {Furlan}},\ }\href
  {https://doi.org/10.1007/BF02785666} {\bibfield  {journal} {\bibinfo
  {journal} {Nuovo Cim. A}\ }\textbf {\bibinfo {volume} {34}},\ \bibinfo
  {pages} {569} (\bibinfo {year} {1976})}\BibitemShut {NoStop}%
\bibitem [{\citenamefont {Jimbo}(1985)}]{Jimbo1985}%
  \BibitemOpen
  \bibfield  {author} {\bibinfo {author} {\bibfnamefont {M.}~\bibnamefont
  {Jimbo}},\ }\href {https://doi.org/10.1007/BF00704588} {\bibfield  {journal}
  {\bibinfo  {journal} {Lett. Math. Phys.}\ }\textbf {\bibinfo {volume} {10}},\
  \bibinfo {pages} {63} (\bibinfo {year} {1985})}\BibitemShut {NoStop}%
\bibitem [{\citenamefont {Klimyk}\ and\ \citenamefont
  {Schm{\"u}dgen}(1997)}]{KlimykSchmudgen1997}%
  \BibitemOpen
  \bibfield  {author} {\bibinfo {author} {\bibfnamefont {A.}~\bibnamefont
  {Klimyk}}\ and\ \bibinfo {author} {\bibfnamefont {K.}~\bibnamefont
  {Schm{\"u}dgen}},\ }\href {https://doi.org/10.1007/978-3-642-60896-4} {\emph
  {\bibinfo {title} {Quantum Groups and Their Representations}}}\ (\bibinfo
  {publisher} {Springer},\ \bibinfo {address} {Berlin},\ \bibinfo {year}
  {1997})\BibitemShut {NoStop}%
\bibitem [{\citenamefont {Liu}\ \emph {et~al.}(2017)\citenamefont {Liu},
  \citenamefont {Nowak},\ and\ \citenamefont {Zahed}}]{LiuNowakZahed2017}%
  \BibitemOpen
  \bibfield  {author} {\bibinfo {author} {\bibfnamefont {Y.}~\bibnamefont
  {Liu}}, \bibinfo {author} {\bibfnamefont {M.~A.}\ \bibnamefont {Nowak}},\
  and\ \bibinfo {author} {\bibfnamefont {I.}~\bibnamefont {Zahed}},\ }\href
  {https://doi.org/10.1016/j.physletb.2017.08.054} {\bibfield  {journal}
  {\bibinfo  {journal} {Phys. Lett. B}\ }\textbf {\bibinfo {volume} {773}},\
  \bibinfo {pages} {647} (\bibinfo {year} {2017})}\BibitemShut {NoStop}%
\bibitem [{\citenamefont {Raz}\ and\ \citenamefont
  {Youssef}(2026)}]{RazYoussef2026}%
  \BibitemOpen
  \bibfield  {author} {\bibinfo {author} {\bibfnamefont {A.}~\bibnamefont
  {Raz}}\ and\ \bibinfo {author} {\bibfnamefont {M.}~\bibnamefont {Youssef}},\
  }\href {https://doi.org/10.1007/JHEP06(2026)215} {\bibfield  {journal}
  {\bibinfo  {journal} {JHEP}\ }\textbf {\bibinfo {volume} {2026}}\bibinfo
  {number} { (06)},\ \bibinfo {pages} {215}}\BibitemShut {NoStop}%
\bibitem [{\citenamefont {Cattaneo}\ and\ \citenamefont
  {Felder}(2000)}]{CattaneoFelder2000}%
  \BibitemOpen
\bibfield  {number} {  }\bibfield  {author} {\bibinfo {author} {\bibfnamefont
  {A.~S.}\ \bibnamefont {Cattaneo}}\ and\ \bibinfo {author} {\bibfnamefont
  {G.}~\bibnamefont {Felder}},\ }\href {https://doi.org/10.1007/s002200000229}
  {\bibfield  {journal} {\bibinfo  {journal} {Commun. Math. Phys.}\ }\textbf
  {\bibinfo {volume} {212}},\ \bibinfo {pages} {591} (\bibinfo {year}
  {2000})}\BibitemShut {NoStop}%
\bibitem [{\citenamefont {Mirzakhani}(2007)}]{Mirzakhani2007}%
  \BibitemOpen
  \bibfield  {author} {\bibinfo {author} {\bibfnamefont {M.}~\bibnamefont
  {Mirzakhani}},\ }\href {https://doi.org/10.1007/s00222-006-0013-2} {\bibfield
   {journal} {\bibinfo  {journal} {Invent. Math.}\ }\textbf {\bibinfo {volume}
  {167}},\ \bibinfo {pages} {179} (\bibinfo {year} {2007})}\BibitemShut
  {NoStop}%
\bibitem [{\citenamefont {Bessis}\ \emph {et~al.}(1980)\citenamefont {Bessis},
  \citenamefont {Itzykson},\ and\ \citenamefont
  {Zuber}}]{BessisItzyksonZuber1980}%
  \BibitemOpen
  \bibfield  {author} {\bibinfo {author} {\bibfnamefont {D.}~\bibnamefont
  {Bessis}}, \bibinfo {author} {\bibfnamefont {C.}~\bibnamefont {Itzykson}},\
  and\ \bibinfo {author} {\bibfnamefont {J.~B.}\ \bibnamefont {Zuber}},\ }\href
  {https://doi.org/10.1016/0196-8858(80)90008-1} {\bibfield  {journal}
  {\bibinfo  {journal} {Adv. Appl. Math.}\ }\textbf {\bibinfo {volume} {1}},\
  \bibinfo {pages} {109} (\bibinfo {year} {1980})}\BibitemShut {NoStop}%
\bibitem [{\citenamefont {Balasubramanian}\ \emph {et~al.}(2023)\citenamefont
  {Balasubramanian}, \citenamefont {Mag{\'a}n},\ and\ \citenamefont
  {Wu}}]{BalasubramanianMaganWu2023}%
  \BibitemOpen
  \bibfield  {author} {\bibinfo {author} {\bibfnamefont {V.}~\bibnamefont
  {Balasubramanian}}, \bibinfo {author} {\bibfnamefont {J.~M.}\ \bibnamefont
  {Mag{\'a}n}},\ and\ \bibinfo {author} {\bibfnamefont {Q.}~\bibnamefont
  {Wu}},\ }\href {https://doi.org/10.1103/PhysRevD.107.126001} {\bibfield
  {journal} {\bibinfo  {journal} {Phys. Rev. D}\ }\textbf {\bibinfo {volume}
  {107}},\ \bibinfo {pages} {126001} (\bibinfo {year} {2023})}\BibitemShut
  {NoStop}%
\bibitem [{\citenamefont {Jensen}(2016)}]{Jensen2016}%
  \BibitemOpen
  \bibfield  {author} {\bibinfo {author} {\bibfnamefont {K.}~\bibnamefont
  {Jensen}},\ }\href {https://doi.org/10.1103/PhysRevLett.117.111601}
  {\bibfield  {journal} {\bibinfo  {journal} {Phys. Rev. Lett.}\ }\textbf
  {\bibinfo {volume} {117}},\ \bibinfo {pages} {111601} (\bibinfo {year}
  {2016})}\BibitemShut {NoStop}%
\bibitem [{\citenamefont {Berkooz}\ \emph {et~al.}(2025)\citenamefont
  {Berkooz}, \citenamefont {Frumkin}, \citenamefont {Mamroud},\ and\
  \citenamefont {Seitz}}]{BerkoozFrumkinMamroudSeitz2025}%
  \BibitemOpen
  \bibfield  {author} {\bibinfo {author} {\bibfnamefont {M.}~\bibnamefont
  {Berkooz}}, \bibinfo {author} {\bibfnamefont {R.}~\bibnamefont {Frumkin}},
  \bibinfo {author} {\bibfnamefont {O.}~\bibnamefont {Mamroud}},\ and\ \bibinfo
  {author} {\bibfnamefont {J.}~\bibnamefont {Seitz}},\ }\href
  {https://doi.org/10.1007/JHEP05(2025)080} {\bibfield  {journal} {\bibinfo
  {journal} {JHEP}\ }\textbf {\bibinfo {volume} {2025}}\bibinfo  {number} {
  (05)},\ \bibinfo {pages} {080}}\BibitemShut {NoStop}%
\bibitem [{\citenamefont {Coghi}\ and\ \citenamefont
  {Garrahan}(2025)}]{CoghiGarrahan2025}%
  \BibitemOpen
\bibfield  {number} {  }\bibfield  {author} {\bibinfo {author} {\bibfnamefont
  {F.}~\bibnamefont {Coghi}}\ and\ \bibinfo {author} {\bibfnamefont {J.~P.}\
  \bibnamefont {Garrahan}},\ }\href {https://doi.org/10.1088/1742-5468/ae099f}
  {\bibfield  {journal} {\bibinfo  {journal} {J. Stat. Mech.}\ }\textbf
  {\bibinfo {volume} {2025}},\ \bibinfo {pages} {103202} (\bibinfo {year}
  {2025})}\BibitemShut {NoStop}%
\bibitem [{\citenamefont {Kac}(1949)}]{Kac1949}%
  \BibitemOpen
  \bibfield  {author} {\bibinfo {author} {\bibfnamefont {M.}~\bibnamefont
  {Kac}},\ }\href {https://doi.org/10.1090/S0002-9947-1949-0027960-X}
  {\bibfield  {journal} {\bibinfo  {journal} {Trans. Amer. Math. Soc.}\
  }\textbf {\bibinfo {volume} {65}},\ \bibinfo {pages} {1} (\bibinfo {year}
  {1949})}\BibitemShut {NoStop}%
\bibitem [{\citenamefont {Lamperti}(1972)}]{Lamperti1972}%
  \BibitemOpen
  \bibfield  {author} {\bibinfo {author} {\bibfnamefont {J.}~\bibnamefont
  {Lamperti}},\ }\href {https://doi.org/10.1007/BF00536091} {\bibfield
  {journal} {\bibinfo  {journal} {Z. Wahrscheinlichkeitstheor. Verw. Geb.}\
  }\textbf {\bibinfo {volume} {22}},\ \bibinfo {pages} {205} (\bibinfo {year}
  {1972})}\BibitemShut {NoStop}%
\bibitem [{\citenamefont {Bagrets}\ \emph {et~al.}(2016)\citenamefont
  {Bagrets}, \citenamefont {Altland},\ and\ \citenamefont
  {Kamenev}}]{BagretsAltlandKamenev2016}%
  \BibitemOpen
  \bibfield  {author} {\bibinfo {author} {\bibfnamefont {D.}~\bibnamefont
  {Bagrets}}, \bibinfo {author} {\bibfnamefont {A.}~\bibnamefont {Altland}},\
  and\ \bibinfo {author} {\bibfnamefont {A.}~\bibnamefont {Kamenev}},\ }\href
  {https://doi.org/10.1016/j.nuclphysb.2016.08.002} {\bibfield  {journal}
  {\bibinfo  {journal} {Nucl. Phys. B}\ }\textbf {\bibinfo {volume} {911}},\
  \bibinfo {pages} {191} (\bibinfo {year} {2016})}\BibitemShut {NoStop}%
\bibitem [{\citenamefont {Watanabe}(2025)}]{Watanabe:2025rwp}%
  \BibitemOpen
  \bibfield  {author} {\bibinfo {author} {\bibfnamefont {M.}~\bibnamefont
  {Watanabe}},\ }\href@noop {} {\bibinfo {title} {A {JT/KPZ} correspondence}}
  (\bibinfo {year} {2025}),\ \Eprint {https://arxiv.org/abs/2511.02529}
  {arXiv:2511.02529 [hep-th]} \BibitemShut {NoStop}%
\bibitem [{\citenamefont {Belokurov}\ and\ \citenamefont
  {Shavgulidze}(2019)}]{BelokurovShavgulidze2019}%
  \BibitemOpen
  \bibfield  {author} {\bibinfo {author} {\bibfnamefont {V.~V.}\ \bibnamefont
  {Belokurov}}\ and\ \bibinfo {author} {\bibfnamefont {E.~T.}\ \bibnamefont
  {Shavgulidze}},\ }\href {https://doi.org/10.1134/S004057791909006X}
  {\bibfield  {journal} {\bibinfo  {journal} {Theor. Math. Phys.}\ }\textbf
  {\bibinfo {volume} {200}},\ \bibinfo {pages} {1324} (\bibinfo {year}
  {2019})}\BibitemShut {NoStop}%
\bibitem [{\citenamefont {Bauerschmidt}\ \emph {et~al.}(2024)\citenamefont
  {Bauerschmidt}, \citenamefont {Losev},\ and\ \citenamefont
  {Wildemann}}]{BauerschmidtLosevWildemann2024}%
  \BibitemOpen
  \bibfield  {author} {\bibinfo {author} {\bibfnamefont {R.}~\bibnamefont
  {Bauerschmidt}}, \bibinfo {author} {\bibfnamefont {I.}~\bibnamefont
  {Losev}},\ and\ \bibinfo {author} {\bibfnamefont {P.}~\bibnamefont
  {Wildemann}},\ }\href@noop {} {\bibinfo {title} {Probabilistic definition of
  the {Schwarzian} field theory}} (\bibinfo {year} {2024}),\ \Eprint
  {https://arxiv.org/abs/2406.17068} {arXiv:2406.17068 [math.PR]} \BibitemShut
  {NoStop}%
\bibitem [{\citenamefont {Stanford}\ and\ \citenamefont
  {Witten}(2017)}]{StanfordWitten2017}%
  \BibitemOpen
  \bibfield  {author} {\bibinfo {author} {\bibfnamefont {D.}~\bibnamefont
  {Stanford}}\ and\ \bibinfo {author} {\bibfnamefont {E.}~\bibnamefont
  {Witten}},\ }\href {https://doi.org/10.1007/JHEP10(2017)008} {\bibfield
  {journal} {\bibinfo  {journal} {JHEP}\ }\textbf {\bibinfo {volume}
  {2017}}\bibinfo  {number} { (10)},\ \bibinfo {pages} {008}}\BibitemShut
  {NoStop}%
\bibitem [{\citenamefont {Nag}\ and\ \citenamefont
  {Sullivan}(1995)}]{NagSullivan1995}%
  \BibitemOpen
\bibfield  {number} {  }\bibfield  {author} {\bibinfo {author} {\bibfnamefont
  {S.}~\bibnamefont {Nag}}\ and\ \bibinfo {author} {\bibfnamefont
  {D.}~\bibnamefont {Sullivan}},\ }\href {https://doi.org/10.18910/8848}
  {\bibfield  {journal} {\bibinfo  {journal} {Osaka J. Math.}\ }\textbf
  {\bibinfo {volume} {32}},\ \bibinfo {pages} {1} (\bibinfo {year}
  {1995})}\BibitemShut {NoStop}%
\bibitem [{\citenamefont {Bargmann}(1947)}]{Bargmann1947}%
  \BibitemOpen
  \bibfield  {author} {\bibinfo {author} {\bibfnamefont {V.}~\bibnamefont
  {Bargmann}},\ }\href {https://doi.org/10.2307/1969129} {\bibfield  {journal}
  {\bibinfo  {journal} {Ann. Math.}\ }\textbf {\bibinfo {volume} {48}},\
  \bibinfo {pages} {568} (\bibinfo {year} {1947})}\BibitemShut {NoStop}%
\bibitem [{\citenamefont {Belavin}\ \emph {et~al.}(1984)\citenamefont
  {Belavin}, \citenamefont {Polyakov},\ and\ \citenamefont
  {Zamolodchikov}}]{BelavinPolyakovZamolodchikov1984}%
  \BibitemOpen
  \bibfield  {author} {\bibinfo {author} {\bibfnamefont {A.~A.}\ \bibnamefont
  {Belavin}}, \bibinfo {author} {\bibfnamefont {A.~M.}\ \bibnamefont
  {Polyakov}},\ and\ \bibinfo {author} {\bibfnamefont {A.~B.}\ \bibnamefont
  {Zamolodchikov}},\ }\href {https://doi.org/10.1016/0550-3213(84)90052-X}
  {\bibfield  {journal} {\bibinfo  {journal} {Nucl. Phys. B}\ }\textbf
  {\bibinfo {volume} {241}},\ \bibinfo {pages} {333} (\bibinfo {year}
  {1984})}\BibitemShut {NoStop}%
\bibitem [{\citenamefont {Bertola}\ \emph {et~al.}(2007)\citenamefont
  {Bertola}, \citenamefont {Corbetta},\ and\ \citenamefont
  {Moschella}}]{BertolaCorbettaMoschella2007}%
  \BibitemOpen
  \bibfield  {author} {\bibinfo {author} {\bibfnamefont {M.}~\bibnamefont
  {Bertola}}, \bibinfo {author} {\bibfnamefont {F.}~\bibnamefont {Corbetta}},\
  and\ \bibinfo {author} {\bibfnamefont {U.}~\bibnamefont {Moschella}},\ }in\
  \href {https://doi.org/10.1007/978-3-7643-7434-1_3} {\emph {\bibinfo
  {booktitle} {Rigorous Quantum Field Theory}}},\ \bibinfo {series} {Progress
  in Mathematics}, Vol.\ \bibinfo {volume} {251}\ (\bibinfo  {publisher}
  {Birkh{\"a}user Basel},\ \bibinfo {year} {2007})\ pp.\ \bibinfo {pages}
  {27--38},\ \Eprint {https://arxiv.org/abs/math-ph/0609080}
  {arXiv:math-ph/0609080} \BibitemShut {NoStop}%
\bibitem [{\citenamefont {Knizhnik}\ and\ \citenamefont
  {Zamolodchikov}(1984)}]{KnizhnikZamolodchikov1984}%
  \BibitemOpen
  \bibfield  {author} {\bibinfo {author} {\bibfnamefont {V.~G.}\ \bibnamefont
  {Knizhnik}}\ and\ \bibinfo {author} {\bibfnamefont {A.~B.}\ \bibnamefont
  {Zamolodchikov}},\ }\href {https://doi.org/10.1016/0550-3213(84)90374-2}
  {\bibfield  {journal} {\bibinfo  {journal} {Nucl. Phys. B}\ }\textbf
  {\bibinfo {volume} {247}},\ \bibinfo {pages} {83} (\bibinfo {year}
  {1984})}\BibitemShut {NoStop}%
\bibitem [{\citenamefont {McRae}(2020)}]{McRae2020}%
  \BibitemOpen
  \bibfield  {author} {\bibinfo {author} {\bibfnamefont {R.}~\bibnamefont
  {McRae}},\ }\href {https://doi.org/10.1016/j.aim.2020.107351} {\bibfield
  {journal} {\bibinfo  {journal} {Adv. Math.}\ }\textbf {\bibinfo {volume}
  {374}},\ \bibinfo {pages} {107351} (\bibinfo {year} {2020})}\BibitemShut
  {NoStop}%
\bibitem [{\citenamefont {Kohno}(1987)}]{Kohno1987}%
  \BibitemOpen
  \bibfield  {author} {\bibinfo {author} {\bibfnamefont {T.}~\bibnamefont
  {Kohno}},\ }\href {https://doi.org/10.5802/aif.1114} {\bibfield  {journal}
  {\bibinfo  {journal} {Ann. Inst. Fourier}\ }\textbf {\bibinfo {volume}
  {37}},\ \bibinfo {pages} {139} (\bibinfo {year} {1987})}\BibitemShut
  {NoStop}%
\bibitem [{\citenamefont {Gaberdiel}(1995)}]{Gaberdiel1994}%
  \BibitemOpen
  \bibfield  {author} {\bibinfo {author} {\bibfnamefont {M.~R.}\ \bibnamefont
  {Gaberdiel}},\ }\href@noop {} {\bibfield  {journal} {\bibinfo  {journal}
  {Commun. Math. Phys.}\ }\textbf {\bibinfo {volume} {173}},\ \bibinfo {pages}
  {357} (\bibinfo {year} {1995})}\BibitemShut {NoStop}%
\end{thebibliography}%

\end{document}